 \documentclass[preprint2]{aastex}                      
\usepackage{natbib}

\begin{document}

\title{The HD\,5980 multiple system: Masses and evolutionary status}

\author{Gloria Koenigsberger}
\affil{Instituto de Ciencias F\'{\i}sicas, Universidad Nacional Aut\'onoma de M\'exico,
Ave. Universidad S/N, Cuernavaca, Morelos, 62210, M\'exico; and Instituto de Astronom\'{\i}a,
Universidad Nacional Aut\'onoma de M\'exico, Apdo. Postal 70-264, M\'exico D.F. 04510, gloria@astro.unam.mx}

\author{Nidia Morrell}
\affil{Las Campanas Observatory, The Carnegie Observatories,
Colina El Pino s/n, Casillas 601, La Serena, Chile, nmorrell@lco.edu}

\author{D. John Hillier}
\affil{Department of Physics and Astronomy, \& Pittsburgh Partile Physis, Astrophysics and Cosmology Center
(PITT PACC), 3941 O'Hara Street, University of Pittsburgh, Pittsburgh, PA 15260, USA, hillier@pitt.edu}

\author{Roberto Gamen}
\affil{Facultad de Ciencias Astron\'omicas y Geof\'{\i}sicas, Universidad Nacional de La Plata,
and Instituto de Astrof\'{\i}sica de La Plata (CCT La Plata-CONICET), Paseo del
Bosque S/N, B1900FWA, La Plata, Argentina, rgamen@fcaglp.unlp.edu.ar}

\author{Fabian R.N. Schneider, Nicol\'as Gonz\'alez-Jim\'enez, Norbert Langer}
\affil{Argelander-Institut f\"ur Astronomie, Auf dem H\"ugel 71, 53121 Bonn, Germany, fschneid@astro.uni-bonn.de, ngonzalez@astro.uni-bonn.de,nlanger@astro.uni-bonn.de}

\and

\author{Rodolfo Barb\'a}
\affil{Departamento de F\'{\i}sica, Av. Juan Cisternas 1200 Norte, Universidad de la Serena,
La Serena, Chile, rbarba@dfuls.cl}

\begin{abstract}
New spectroscopic observations of the LBV/WR multiple system HD\,5980 in the Small Magellanic Cloud are
used to address the question of the masses and evolutionary status of the two very luminous stars in the 
19.3d eclipsing binary system. Two distinct 
components of the \ion{N}{5} 4944 \AA\ line are detected in emission and their radial velocity 
variations are used to derive masses of 61 and 66 M$_\odot$, under the assumption that binary
interaction effects on this atomic transition are negligible.  We propose that this binary system
is the product of quasi-chemically homogeneous evolution with little or no mass transfer. Thus, 
both of these binary stars may be candidates for gamma-ray burst progenitors or even pair instability 
supernovae. Analysis of the photospheric absorption lines belonging to the third-light object in the system 
confirm that it consists of an O-type star in a 96.56d eccentric orbit (e$=$0.82) around an unseen companion. 
The 5:1 period ratio and high eccentricities of the two binaries suggest that they may constitute 
a hierarchical quadruple system. 

\end{abstract}


\section{Introduction}

Massive stars play an important role in the chemical evolution of galaxies
and in shaping the interstellar medium. They are found in regions of recent
star formation and near the Galactic center, and they are associated with
fascinating phenomena such as the formation of neutron stars and black holes.
But despite their importance,  there are still gaps in our understanding of
their structure and evolution.
One of these gaps involves the manner in which they reach the final
stages of evolution and explode as supernovae (SN).   In particular, there
is great uncertainty concerning the amount of mass that is lost in the post-Main
Sequence evolutionary stages, thus leading to uncertainties in the stellar structure
prior to the final SN event. Central to this issue is the role played by the
Luminous Blue Variables (LBV).

According to current evolutionary scenarios, classical LBVs are stars in transition from the Main Sequence
to the Wolf-Rayet stage and they are in the process of shedding their outer layers to expose the
products of the CNO nuclear reactions taking place in their centers (\citet{1987A&A...182..243M};
\citet{1994A&A...290..819L}; \citet{2012ARA&A..50..107L}).
Enhanced mass-loss rates are observed to occur during the maxima of  
the variability  cycles when up to $\sim$10 M$_\odot$ can be ejected, 
as illustrated by $\eta$ Carinae's mid-19th century event (\citet{1994PASP..106.1025H}).
However, the underlying processes that 
cause the LBV instability  are unknown (see, for example, \cite{2012ASPC..465..207V}).  Thus, a crucial ingredient 
in the models of  massive star evolution is missing.  This problem is compounded by the fact that
the current mass of the LBVs is in general poorly constrained.  

The Small Magellanic Cloud harbors  an LBV eclipsing binary 
that has been studied since the 1950's and which underwent a major eruption
in 1994.  It has  the highest luminosity of all the well-studied LBVs, except perhaps for $\eta$
Carinae,  and also shares with $\eta$ Car the  exceptional characteristic of showing  
a Wolf-Rayet type spectrum. Furthermore, both systems are associated with  massive and young
stellar clusters, HD\,5980 lying in the periphery of NGC 346. Its wind properties and many of its 
fundamental parameters have been derived in a series of studies based on {\it IUE}, {\it HST/STIS} 
and optical observations (\citet{2009A&A...503..963P}, \citet{2011AJ....142..191G}, \citet{2008RMxAA..44....3F},
\citet{2004RMxAA..40..107K}, \citet{2010AJ....139.2600K}).  The mass determinations so for, however, have
a large degree of uncertainty, a problem we address in this investigation.  In addition, 
the long baseline of data now available allows a re-assessment of the nature of the third 
component of the system, which we now can confirm consists of a binary system. 

The  organization of this paper is as follows: In Section 2 we review the characteristics
of the HD\,5980 system; in Sections 3 and 4 we describe, respectively, the observational data and 
the radial velocity measurements; Section 5 presents the oribital solutions; in Section 6 we 
discuss the evolutionary paths that may explain HD\,5980's current state; and 
in Section 7 we summarize the conclusions.

\section{Description of the HD\,5980 system}

The spectrum of HD\,5980 is the sum of several components.  Most prominent are the  two emission-line
stars belonging to the close eclipsing binary system (P$_{AB}$=19.3 d). Following the convention introduced
by \citet{1996RMxAC...5...85B},  the star ``in front" at orbital phase $\phi=$0.00 is labelled {\it Star A}
and its companion is labelled {\it Star B}.  In addition, there is a third component, {\it Star C},  
whose photospheric absorptions remain stationary on the 19.3d orbital timescale and which contributes a 
significant fraction of the observed light (\citet{1991IAUS..143..229B, 1988ASPC....1..381N,  
2002ApJ...581..598K,  2009A&A...503..963P}).

HD\,5980's emission-line  spectrum is highly  variable on orbital and decade-long timescales.
The short-timescale variability is produced by the periodic Doppler motion of the emission lines
that are formed in the winds of {\it Stars A} and {\it B} combined with binary interaction
effects, among which the ones that are most likely dominant are irradiation, wind-wind collisions and
wind eclipses. Different lines are affected to a different degree, depending on their transition
probabilities and formation mechanisms.  The long-term variation is due to {\it Star A}'s changing 
wind structure, which appears to be triggered by  variations in the hydrostatic radius 
(\citet{1998ApJ...496..934K,  2011AJ....142..191G}).

{\it Star B} is thought to be a WN4 star (\citet{1982ApJ...257..116B,1988ASPC....1..381N}),
and hence it has a  stellar wind that contributes significantly to the emission-line spectrum.
Traces of its broad UV  emission lines can be seen underlying the narrow emissions 
that appeared during the 1994 eruptive event (see \citet{2004RMxAA..40..107K}, Fig. 11).
Furthermore, the shape of the eclipse light curve observed in the late 1970s leads to the
conclusion that {\it Star B} possesses an extended occulting region having a radius of 0.269$a$,
where $a$ is the semi-major axis of the orbit (\citet{2009A&A...503..963P}).  Both of these facts 
lend credence supporting the view that {\it Star B} is a WNE-type star. 

The spectrum of {\it Star C} is that of an O-type supergiant. Its optical spectrum 
is dominated by H-Balmer, He I and other low-ionization photospheric absorption lines, while 
in the UV it presents a modest P Cygni Si IV 1400 \AA\ profile (\citet{2011AJ....142..191G}).  
The radial velocity of the O III 5592 \AA\ photospheric line was found to display a periodic 
behavior with P$_C \sim$97d,  and the solution to this RV curve yielded a highly eccentric 
e$\sim$0.8 orbit, leading to the conclusion that {\it Star C} is also a binary system 
(\citet{2000PhDT.........2S}).   

The determination of the masses of the individual components in the system is a challenging  
problem due to the very broad and variable emission lines.  Using photographic spectra  obtained 
during 1981-1983, Niemela (1988) estimated masses  M$_A\sim$46 M$_\odot$ and  M$_B\sim$43 M$_\odot$ 
for {\it Star A} and {\it B}, respectively.
\citet{2008RMxAA..44....3F} analyzed high dispersion spectra obtained in 1998, 1999 and 2005 and concluded 
that, if the observed line profiles can be interpreted as being formed primarily in the winds of 
{\it Star A} and {\it Star B} (as opposed to in the wind-wind collision region; see Section 5.3),
then the measured RV variations implied M$_A\sim$ 58--79 M$_\odot$ and M$_B\sim$ 51--67 M$_\odot$.
The stellar wind analysis performed by \citet{2011AJ....142..191G} led to the conclusion that  
for a stable solution connecting the photosphere with the stellar wind to be possible, 
M$_A \geq$90 M$_\odot$.
\citet{1998ApJ...497..896M} assumed that at least part of the photospheric absorptions in the 
spectrum arise in {\it Star A}, from which they derived  masses M$_A\sim$28--38 M$_\odot$ and  
M$_B\sim$6--18 M$_\odot$.  
A possible explanation for the  modulation  on the 19.3d period of the 
absorption lines claimed by Moffat et al. (1998) was put forth by \citet{2004IAUS..215..111G}.
They showed that if {\it Star A} is a rapid rotator, the superposition of its broad and shallow 
photospheric lines could  produce a slight RV fluctuation in {\it Star C}'s sharper lines and 
lead to the false conclusion that they follow the 19.3d orbital period, albeit, with a very small
RV curve amplitude.
 
\section{Observational material}

\begin{figure}
\plotone{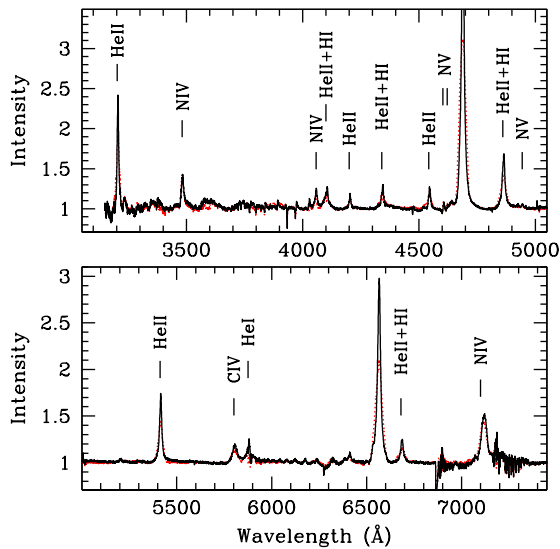}
\caption{Overview of the optical spectrum in 2009 at orbital phases 0.038  and
0.944 (dots). At $\phi$=0.038,  most emission lines have larger
line-to-continuum ratios due in part to the eclipse of {\it Star B} by {\it Star A}.
\label{fig_lowres_whole}}
\end{figure}

The observations of HD\,5980 analyzed in this paper were obtained at Las Campanas Observatory and 
consist of 7 low dispersion spectra obtained with the 6.5m Magellan telescopes in 2009 and 2010; 
25 echelle spectra obtained with the du Pont telescope between 2006 and 2013; and 6  spectra obtained 
on the 6.5 Clay Magellan telescope between 2007 and 2012. Tables \ref{table_lowres}  and \ref{table_highres} 
list the year the spectrum was obtained, the telescope and instruments used, 
Heliocentric Julian Date of observation, orbital phase for P=19.2654 d and T$_0$= HJD 2443158.705 
(\citet{1997A&A...328..269S}).   

\begin{deluxetable}{lllll}
\tablecaption{Low dispersion LCO spectra \label{table_lowres}}
\tablecolumns{5}
\tablewidth{0pt}
\tablehead{
\colhead{Year} & \colhead{Telescope} & \colhead{Instrument} &\colhead{HJD} &\colhead{Phase}\\
\colhead{} & \colhead{} & \colhead{} & \colhead{-2450000} &\colhead{19.26d}
}
\startdata
2009    &Baade Mag 1 &IMACS    &4900.496 &0.476          \\
2009    &Baade Mag 1 &IMACS    &4902.494 &0.579           \\
2009    &Baade Mag 1 &IMACS    &5072.798 &0.419          \\
2009    &Clay  Mag 2 &MagE     &5140.702 &0.944          \\
2009    &Clay  Mag 2 &MagE     &5142.508 &0.038          \\
2009    &Clay  Mag 2 &MagE     &5143.504 &0.089          \\
2010    &Clay  Mag 2 &MagE     &5527.770 &0.035         \\
\enddata
\end{deluxetable}

The Baade 6.5m (Magellan-I) low dispersion spectra were obtained with IMACS in its long camera 
mode (f4) with a 600 line mm$^{-1}$ grating that provided spectral coverage from roughly 3700 to 6800 \AA\ 
at a spectral resolution of 1.4 to 1.5 A (R $\sim$ 4000).  
One of these spectra was described in \citet{2010AJ....139.2600K}.
The Clay 6.5m (Magellan-II) low dispersion spectra  were obtained with the  Magellan Echellette (MagE) 
using the 1\arcsec ~slit  providing a spectral resolution of 1 \AA.  Thirteen  echelle orders were extracted 
covering the wavelength region from 3130 A to 9400 \AA. The signal-to-noise ratio ranges from 100 to 200 
for a single 150 s exposure. The usual ThAr comparison lamp was used for wavelength calibration,  
and the data were reduced using an adaptation of the {\it mtools} package, available for the reduction
of MIKE spectra (see below) in combination with {\it IRAF}\footnote{{\it IRAF} is written
and supported by the National Optical Astronomy Observatories (NOAO) in Tucson, Arizona. NOAO is
operated by the Association of Universities for Research in Astronomy (AURA), Inc. under cooperative
agreement with the National Science Foundation.} echelle  routines for the reduction of MIKE 
spectra\footnote{{\it mtools} package;
http://www.lco.cl/telescopes-information/magellan/instruments/mike/iraf-tools}.
Spectra of spectrophotometric standard stars observed during the same
nights of observation were used to derive a sensitivity function.
The individual flux calibrated echelle orders were then normalized and
merged in the final spectrum. One of these MagE observations was already described in
\citet{2011AJ....142..191G}. 

The high resolution du Pont telescope spectra were obtained with the
echelle spectrograph and a 1\arcsec ~slit. The spectral resolution of these
data ranges from 0.15 to 0.22~\AA\ (R $\sim$ 25000) and the wavelength
coverage goes from 3500 to 8800 \AA.

The high resolution Clay (Magellan-II) spectra were obtained with
the Magellan Inamori Kyocera Echelle spectrograph (MIKE) using a
0.7\arcsec ~slit and applying a 2$\times$2 binning to both blue and red
detectors.  This configuration results in a spectral resolution
of $\sim$34000 (FWHM ranging from 0.10 to 0.25 \AA).
Reductions were carried out with the specially designed {\it IRAF}
scripts contained in the {\it mtools} package
developed by Jack Baldwin and available at the Las Campanas web site.
The reduced spectra were flux calibrated using spectra of
spectrophotometric standards obtained during the same observing nights.

In addition to the new spectra described above, we used data obtained in 1998, 1999,
2005 and 2006 with  FEROS  on the 2.2m MPI telescope at La Silla Observatory, previously 
described in \citet{2002ASPC..260..489K}, Schweickhardt (2000) and Foellmi et al. (2008).

\begin{figure}
\plotone{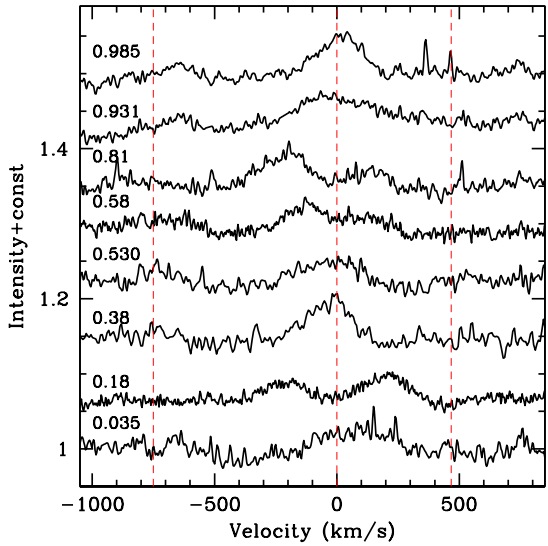}
\caption{Montage of selected NV 4944 line profiles from the LCO echelle data covering the
19.3d orbital cycle, plotted on
a velocity scale corrected for an assumed local SMC velocity of $+$150 km s$^{-1}$.  {\it Star A}  occults
{\it Star B} at $\phi$=0, and approaches the observer at $\phi$=0.81.
Orbital phases with two digits only after the decimal point indicate spectra that have been averaged
over phase.  Spectra are shifted vertically for clarity.   The dotted lines indicate the rest
wavelength location of the  weaker neighboring \ion{N}{5} lines.
\label{fig_nv_LCO}}
\end{figure}

\begin{figure}
\plotone{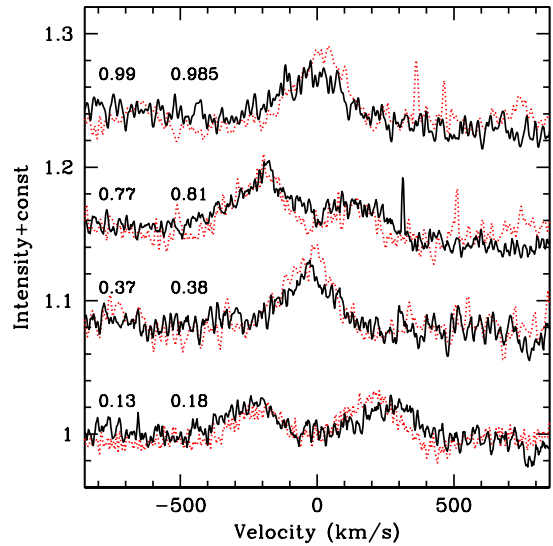}
\caption{Montage of the \ion{N}{5} 4944 \AA\ line in the 2005-2006 FEROS spectra showing the same
behavior as oberved in the 2009-2011 LCO data.  The dotted spectra are the same LCO spectra as
plotted in Fig.~\ref{fig_nv_LCO}.
The orbital phases for FEROS (left) and LCO (right) are indicated. The velocity scale
is corrected for the adopted systemic SMC velocity of $+$150 km s$^{-1}$.
\label{fig_nv_FEROS} }
\end{figure}

\section{Radial velocities and orbital solutions}

In the following sections we will describe the procedure followed for obtaining the RV curves 
of the two emission-line objects and of the star responsible for the absorption lines. 
For this, it is useful to keep in mind that the eclipses of the 19.3d binary occur at $\phi$=0.00 ({\it Star A} 
``in front'') and $\phi$=0.36  ({\it Star B} ``in front'') and maximum receding velocity of 
{\it Star A} is at $\phi\sim$0.1.

\subsection{Emission lines}

Fig.~\ref{fig_lowres_whole} shows the spectra obtained in 2009 at orbital phases $\phi$=0.038 and 0.944 
in the $\lambda\lambda$3100--7400 \AA\ wavelength region. The dominant features are 
emission lines of \ion{He}{2}, \ion{N}{4}, \ion{N}{5}, and \ion{C}{4};  He I is weakly present in emission at $\lambda$5875.   

All the major emission lines undergo prominent shape changes over the orbital cycle, similar to those
found in previous optical data sets (\citet{1982ApJ...257..116B, 2000A&A...361..231B},  Moffat 
et al. 1998; Foellmi et al. 2008; Koenigsberger et al. 2010).  The dominant type of variability involves the
full width at half maximum intensity (FWHM) which systematically is greater at elongations than at
eclipses.    {\it Star A} and {\it Star B} are believed to 
possess stellar winds of similar ionization characteristics and the orbital motion of these stars
combined with the interactions and the eclipses  produces the FWHM variations.  In Section 5.3 we discuss
the validity of this assumption in the light of the various interaction effects that are present
in the {\it Star A} + {\it Star B} system, and we argue in favor of the possibility of using the weaker 
high ionization emission lines for deriving the orbital parameters.  Because the ionization rapidly
decreases outward, the bulk of their emission arises 
in the innermost regions of the wind and are thus less sensitive to external perturbations. 

Narrow lines are  also the best ones to measure on the echelle spectra, where uncertainties in the 
rectification of the individual orders represents a challenge for accurate measurement of very broad and 
variable emissions. In addition, it is advisable to select lines that lie near the
center of the order where the S/N is greatest and that are not blended with other lines.  We found that
the line that is best suited for the RV curve analysis is \ion{N}{5} 4944 \AA.  This is one of the
three  NV lines that lie in the same echelle order at $\lambda\lambda$ 4944.37, 4932.00 and 4952.07 \AA.
The first of these lies closest to the center of the order in most of our spectra and it is the
only one of these three lines that is clearly visible. This is consistent with its intensity in our
{\it Star A} CMFGEN model, where it has an equivalent width $\sim$8 times greater than that of 
$\lambda$4932 and 4952 \AA.  

Fig.~\ref{fig_nv_LCO}  displays the rectified\footnote{The echelle orders were rectified by tracing
the continuum level on each order and then using the resulting curve for the normalization.} echelle 
orders containing these \ion{N}{5} transitions and stacked from bottom to top with increasing orbital phase 
plotted on a velocity scale corrected for an assumed SMC systemic velocity of +150 km s$^{-1}$. 
It  shows that \ion{N}{5} 4944 \AA\ splits into two components during the phase intervals:
0.09--0.24 (when {\it Star A} is receding from the observer) and 0.58--0.81 (when it is approaching).
A similar pattern is also observed to occur in the FEROS data of 2005-2006, as 
illustrated in Fig. ~\ref{fig_nv_FEROS}.

The RVs of  \ion{N}{5} 4944 were  measured with a two-Gaussian deblending routine in {\it IRAF}.
These measurements were first performed on the non-rectified LCO echelle orders and then repeated 
on the rectified orders. In the spectra in which the two components are well separated,
the largest uncertainty is $\sim$10 km s$^{-1}$ and stems from the definition of the continuum level.  
At eclipse phases where only one centrally-located emission is visible, the measured RV is assigned 
to the uneclipsed component, with the uncertainty stemming also from the continuum level.
When the lines are not well separated, the uncertainty can be as large as $\sim$30 km s$^{-1}$.  

\begin{figure}
\plotone{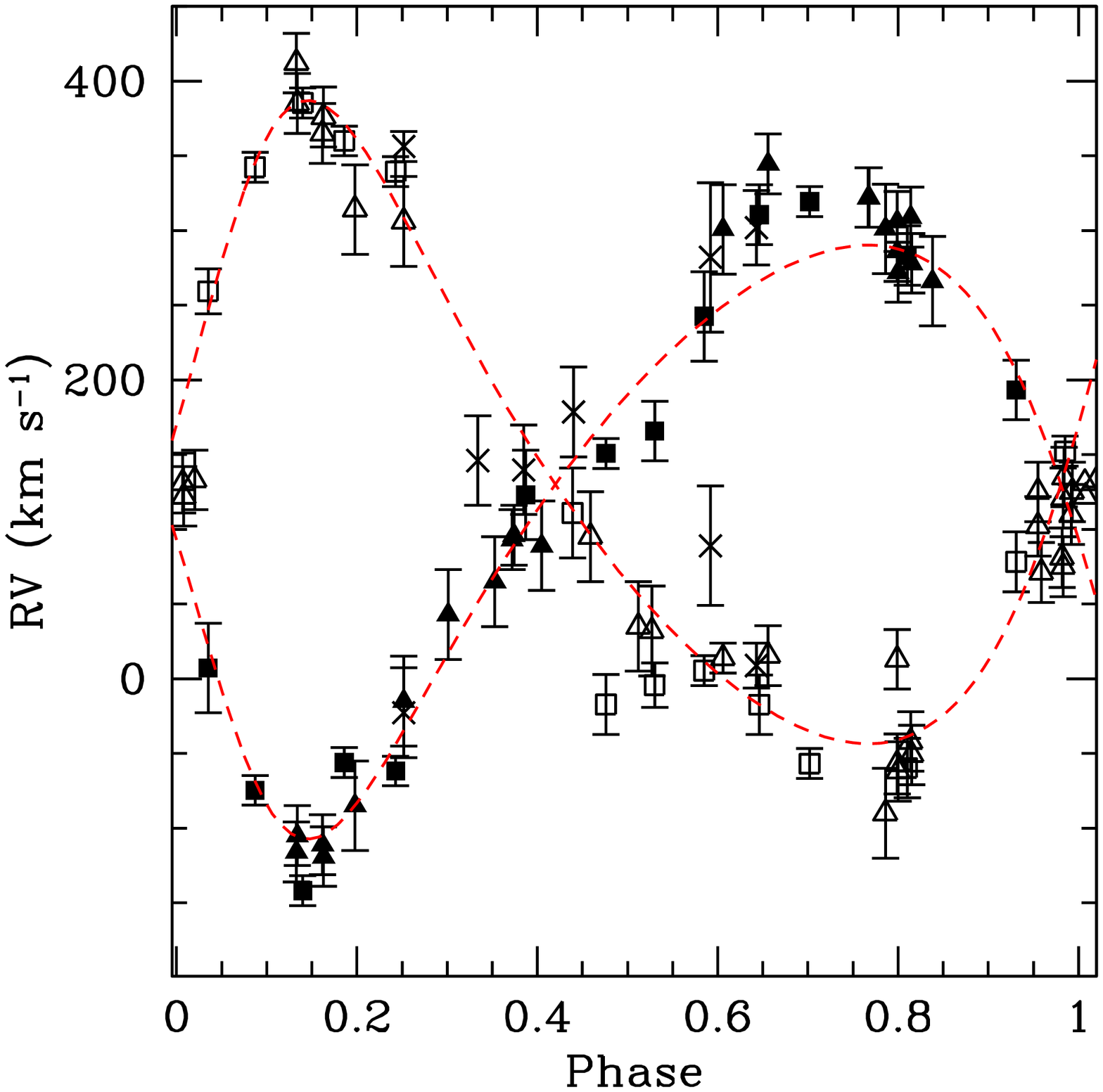}
\caption{Radial velocities of the NV 4944 \AA\ emission line  obtained in 1999--2008 (triangles),
2009--2012 (squares) and 2013 (crosses) plotted as a function of the 19.2654d orbital phase. Error bars
indicate the measurement uncertainties.  The dash curves correspond to the orbital solution
listed in Table \ref{table_RV_LC} with a systematic shift of -20 km s$^{-1}$.
\label{fig_RVs_AandB}}
\end{figure}

Consistent measurements were not possible on a number of spectra due in general to a combination of 
too poor S/N, spurious features and cosmic ray hits, and these spectra have been excluded from the 
\ion{N}{5} 4944 analysis.  The 2005--2006 FEROS exposures were performed in pairs, and each individual
spectrum was measured and the results averaged.
 The results are presented in Table ~\ref{table_NV} and Fig.~\ref{fig_RVs_AandB}.

\small{
\begin{deluxetable}{lllll}
\tablecaption{High dispersion LCO spectra \label{table_highres}}    
\tablecolumns{5}                                                                       
\tablewidth{0pt}
\tablehead{
\colhead{Year}&\colhead{Telescope}&\colhead{Instrument}&\colhead{HJD}&\colhead{Phase } \\
\colhead{}    & \colhead{}       & \colhead{}          & \colhead{-2450000} &\colhead{19.2654d}
}
\startdata
2006                          &DuPont      &echelle        &3920.861 &0.626           \\
2007                          &Clay  Mag 2 &MIKE           &4342.731 &0.524          \\   
2008                          &DuPont      &echelle        &4670.799 &0.553           \\
2008                          &DuPont      &echelle        &4671.823 &0.606           \\
2008                          &DuPont      &echelle        &4672.799 &0.656           \\
2008                          &Clay  Mag 2 &MIKE           &4774.800 &0.951         \\ 
2009                          &Clay  Mag 2 &MIKE           &5161.687 &0.033          \\            
2010                          &Clay  Mag 2 &MIKE           &5205.520 &0.308          \\            
2010                          &DuPont      &echelle        &5341.902 &0.387          \\
2010                          &DuPont      &echelle        &5342.889 &0.439        \\
2010                          &DuPont      &echelle        &5480.501 &0.581        \\           
2010                          &DuPont      &echelle        &5481.756 &0.646         \\
2010                          &DuPont      &echelle        &5482.811 &0.702          \\           
2010                          &DuPont      &echelle        &5506.500 &0.931          \\
2010                          &DuPont      &echelle        &5507.542 &0.985         \\           
2010                          &DuPont      &echelle        &5508.508 &0.035           \\           
2010                          &DuPont      &echelle        &5509.500 &0.087          \\
2010                          &DuPont      &echelle        &5510.521 &0.140           \\              
2010                          &DuPont      &echelle        &5511.546 &0.193          \\
2010                          &DuPont      &echelle        &5512.502 &0.243          \\            
2011                          &Clay  Mag 2 &MIKE           &5761.866 &0.186         \\
2011                          &DuPont      &echelle        &5844.530 &0.477         \\
2011                          &DuPont      &echelle        &5845.564 &0.530         \\
2011                          &DuPont      &echelle        &5846.620 &0.585         \\
2012                          &Clay  Mag 2 &MIKE           &6084.887 &0.953          \\            
2012                          &DuPont      &echelle        &6120.766  &0.815         \\
2013                          &DuPont      &echelle        &6496.804  &0.334         \\
2013                          &DuPont      &echelle        &6497.804  &0.385         \\
2013                          &DuPont      &echelle        &6498.833  &0.440         \\
2013                          &DuPont      &echelle        &6501.750  &0.592         \\
2013                          &DuPont      &echelle        &6502.754  &0.643         \\
2013                          &DuPont      &echelle        &6533.737  &0.252         \\
\enddata
\end{deluxetable}
}             

We searched for other NV lines that could be measured in a similar fashion as NV 4944,  but 
they are all either too weak or lie too close to the edges of the orders, except for 
$\lambda\lambda$ 4603--19 \AA\ which, however, possess strong overlapping P Cygni absorption 
components which do not allow reliable measurements.

\begin{figure}
\plotone{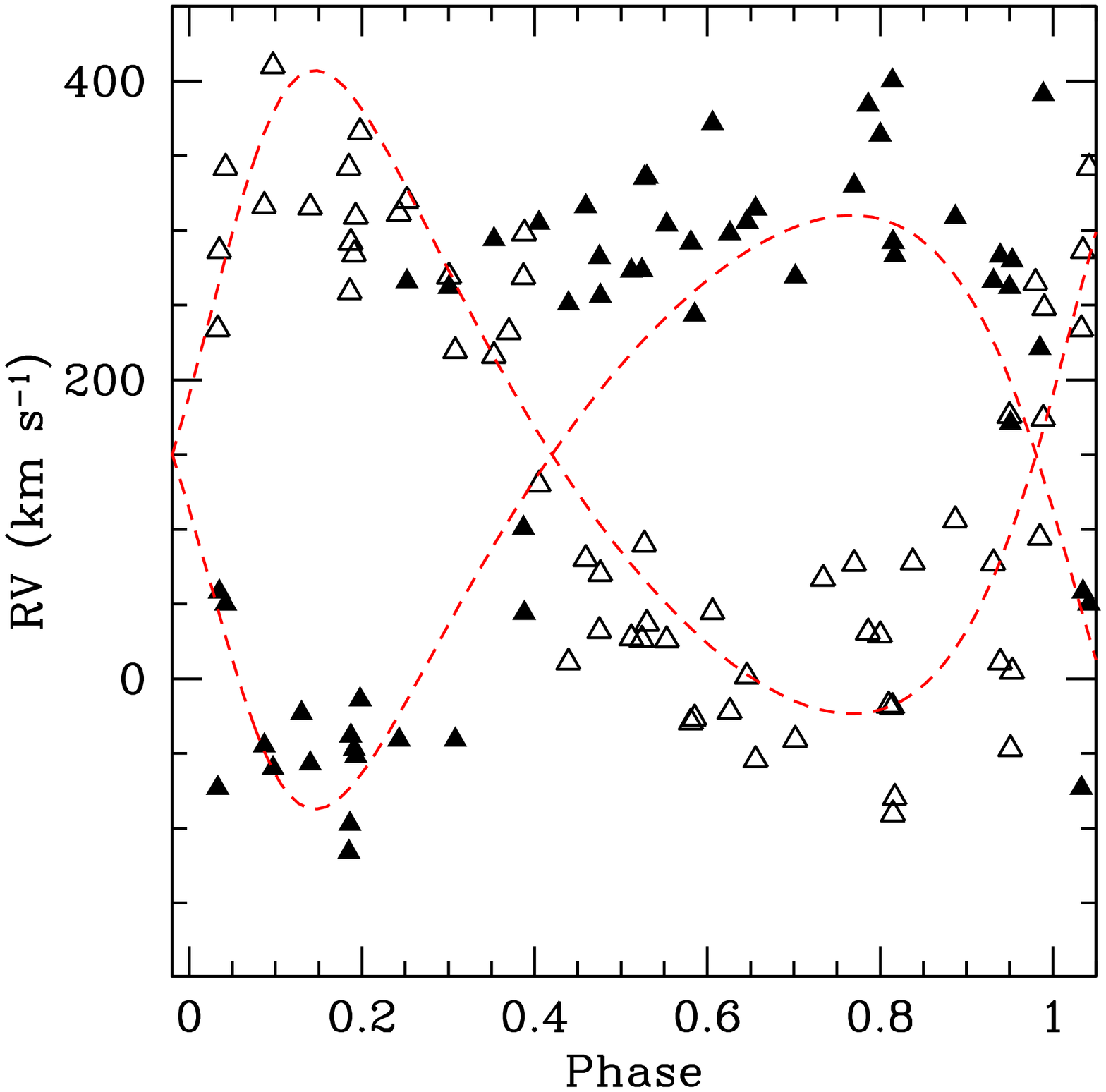}
\caption{Radial velocity  measurements of the NIV~4057 emission line from the rectified LCO orders and
from the archival FEROS data plotted as a function of the 19.2654 d orbital phase.  Open and filed-in
symbols correspond, respectively, to {\it Star A} and {\it Star B}.  The larger dispersion here
compared to the plot of \ion{N}{5} 4944 \AA\ is most likely related to stronger
interaction effects in this \ion{N}{4} emission line.
\label{fig_RVs_NIV4057}}
\end{figure}

Another line that is amenable for measurement is \ion{N}{4} 4057 \AA. It was measured by Foellmi et al. (2008) 
in the 1999 and 2005 FEROS spectra by assuming that it consists of two components that are 
Doppler-shifted due to orbital motion.  We have
applied the same procedure to measure this line in the LCO spectra listed in Table ~\ref{table_highres}.  
To do so, we fixed the limits for the deblending routine at $\pm$500 km/s, which therefore neglects
the broad emission-line wings.  The resulting RVs of the LCO and FEROS data sets are plotted in 
Fig ~\ref{fig_RVs_NIV4057}, which shows that the RV variations of this line are similar to those
found for \ion{N}{5} 4944 \AA.  There is, however, a systematic  difference in the LCO and FEROS RV curves 
of \ion{N}{4} 4057 \AA\ at orbital phases 0.25--0.45,   most likely a consequence of epoch-dependent variations 
in {\it Star A}'s wind structure.\footnote{In this phase interval {\it Star B} is 
``in front" of the companion, and changes in the wind structure of {\it Star A} would affect the 
geometry of the wind-wind collision region; however, given the larger {\it Star A}  wind density,
the irradiation effects would also be more important and back-scattered emission would contribute to the
``blue'' emission-line wing.}

\begin{deluxetable}{llrrllrr}
\setlength{\tabcolsep}{0.005in}
\tablecaption{Radial velocity measurements of NV 4944 \label{table_NV}}
\tablecolumns{8}
\tablewidth{0pt}
\tablehead{
\colhead{HJD$^a$} &\colhead{Phase$^b$}& \colhead{A$^c$} & \colhead{B$^c$}  &\colhead{HJD$^a$} &\colhead{Phase$^b$}& \colhead{A$^c$} & \colhead{B$^c$}  
}
\startdata
   {\bf LCO}&          &        &            & {\bf FEROS}        &       &          &       \\
    4671.823&     0.606&     14 &    301     &  1375.922&     0.527&     32 &   \nodata   \\
    4672.799&     0.656&     15 &    344     &  1380.898&     0.786&  $-$90 &    301      \\
    5341.897&     0.380&   \nodata&  123     &  1381.902&     0.838&  \nodata&    266    \\
    5342.889&     0.439&    111 & \nodata    &  1388.840&     0.198&    314 &  $-$85      \\
    5481.756&     0.646&  $-$17 &    311     &  1389.883&     0.252&    306 &  $-$15      \\
    5482.811&     0.702&  $-$57 &    319     &  1390.828&     0.301&    160 &     43      \\
    5506.500&     0.931&     78 &    193     &  1391.832&     0.353&   \nodata &    65   \\
    5507.542&     0.985&    152 &   \nodata  &  1392.828&     0.405&   \nodata &    89  \\
    5508.508&     0.035&    259 &     7      &  1393.871&     0.459&   \nodata&     95 \\
    5509.500&     0.087&    343 &  $-$75     &  1394.891&     0.512&     35 &   \nodata \\
    5510.521&     0.140&    386 & $-$142     &  3538.885&     0.799&     13 &    306 \\
    5512.502&     0.243&    339 &  $-$61     &  3541.894&     0.955&    113 &   \nodata \\
    5761.866&     0.186&    360 &  $-$56     &  3561.877&     0.992&    117 &   \nodata \\
    5844.530&     0.477&  $-$17 &    151     &  3576.798&     0.767&  $-$98 &    322 \\
    5845.564&     0.530&   $-$4 &    166     &  3635.510&     0.814&  $-$47 &    294 \\
    5846.620&     0.585&      6 &    242     &  3638.763&     0.983&    128 &   \nodata \\
    6120.767&     0.810&  $-$60 &    283     &  3641.663&     0.133&    399 & $-$110 \\
    6496.804&     0.334&  \nodata&   146     &  3665.535&     0.373&   \nodata&     94 \\
    6497.804&     0.385&  \nodata&   139     &  3715.798&     0.982&     78 &   \nodata \\
    6498.833&     0.440&  \nodata&   178     &  3716.536&     0.020&    133 &   \nodata\\
    6501.750&     0.592&     89 &    282     &  3731.546&     0.799&  $-$59 &    289 \\
    6502.754&     0.643&      9&     302     &  3734.615&     0.959&     71 &   \nodata\\
    6533.737&     0.252&   356  & $-$22      &  3735.553&     0.007&    126 &   \nodata \\
            &          &        &            &  3738.544&     0.162&    369 & $-$115 \\
\enddata
\end{deluxetable}

\subsection{Photospheric absorption lines}

It is now well-established that the sharp photospheric absorptions (see Fig.~\ref{fig_lowres_whole})
 arise in {\it Star C}, but the presence of such absorptions arising in {\it Star A} and/or {\it Star B} 
cannot be excluded.  A search for photospheric absorptions has repeatedly been conducted
(Niemela 1988; Koenigsberger et al. 2002) with no success. This can be explained
if the optical depth of the stellar wind is too large for the spectrum of the hydrostatic radius
to be visible, an explanation that is supported by the CMFGEN model which indicates that 
the optical depth at the sonic radius is $\sim$1.8 (\citet{2011AJ....142..191G}; model 
fit to the 2009 spectra).  However, there is evidence that the wind is becoming weaker so that  
photospheric absorptions may be expected to become visible,  prompting a new search in the current data set.

Ten of the du Pont Telescope observations were obtained over a 32 night timespan in 2010 and cover
the orbital phases 0.58--1.24.  We first searched these spectra for photospheric absorption lines
with RV variations that would allow them to be  associated with either {\it Star A} or {\it Star B}.
Although many potential identifications were made, we were unable to convincingly find  any one feature
that appears in several spectra with RV shifts that would prove that it arises in one of these stars.
The search was then extended to the other spectra with similar inconclusive results.
It is worth noting that photospheric absorptions in a rapidly rotating star are difficult to
detect in general and more so if they are superposed on broad and variable emission lines as might
be the case for {\it Star A} and/or  {\it Star B}. 

\begin{figure} [!h, !t, !b]
\includegraphics[width=0.48\linewidth]{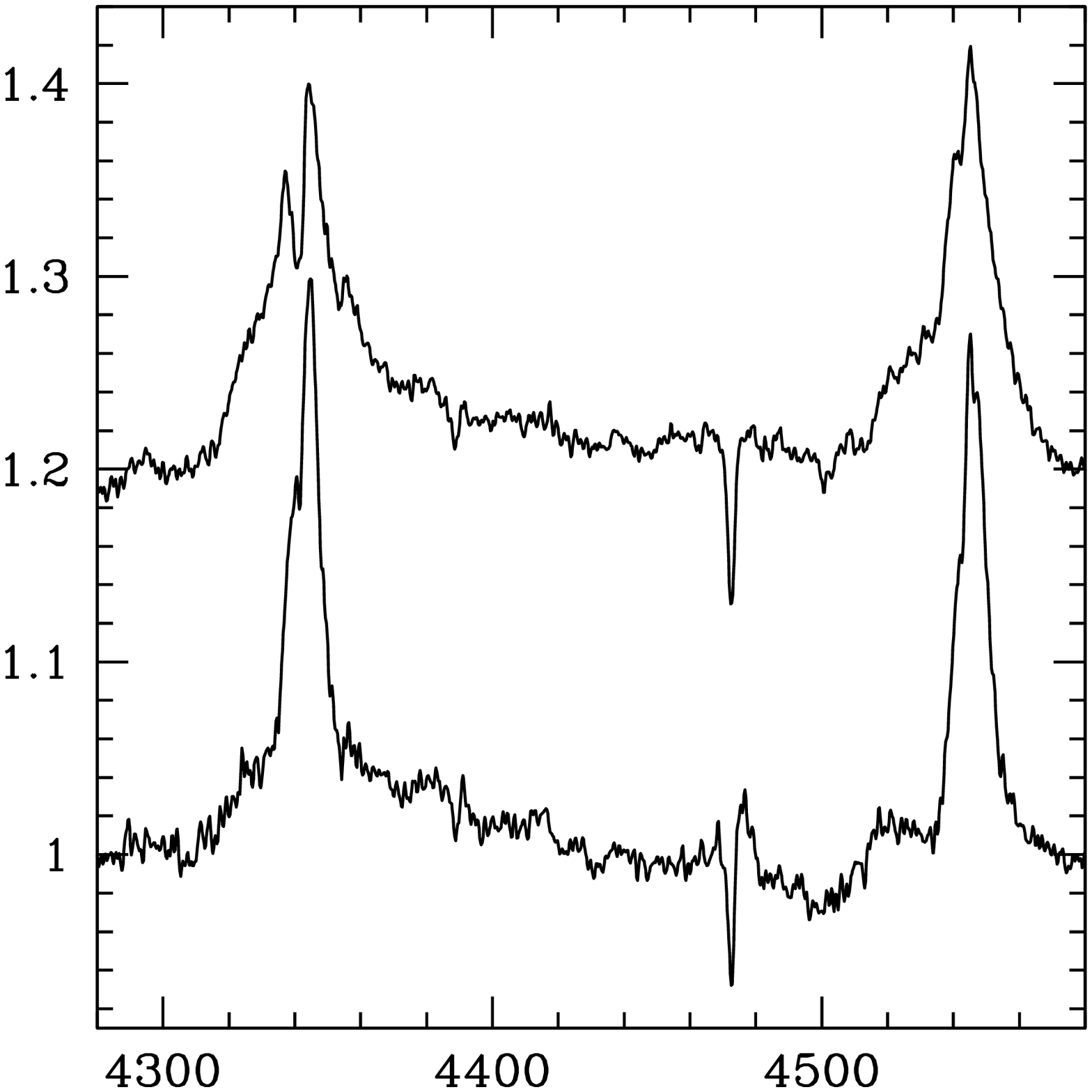}
\includegraphics[width=0.48\linewidth]{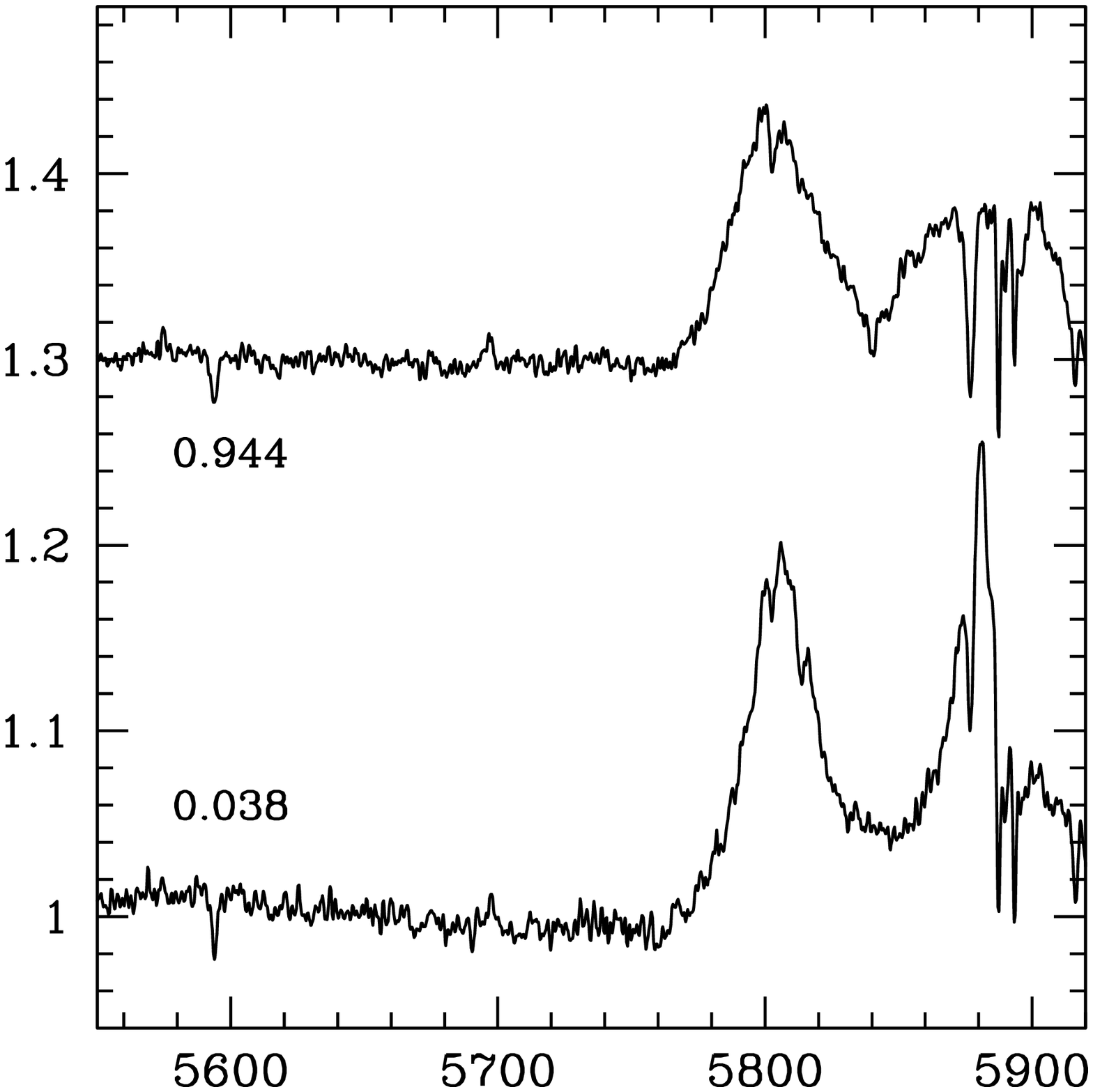}
\caption{Selected spectral regions showing photospheric absorption from {\it Star C}
in the low dispersion spectra obtained at orbital phases $\phi_{AB}$=0.038 (lower tracing) 
and 0.944 (upper tracing; displaced vertically for clarity).  Shown are He I 4387 
and 4471 \AA\ (left panel)  and O III 5592, \ion{C}{4} 5801, 5812 and He I 5875 \AA\ (right panel). 
\label{fig_lowres_fragments}}
\end{figure}

We now shift our focus to the photospheric absorptions that are visible in the spectra
and do not partake in the 19.3d orbit. These include   
the H-Balmer series out to H20, as well as prominent He I 4471, 5875 \AA, O III 5592 \AA\ and \ion{C}{4} 5806, 5812 \AA, 
and weaker absorption attributable to \ion{He}{2} 4542 and 5411 \AA. The  absorptions 
are most clearly visible at orbital phases when the underlying emissions are broadest 
(such as at $\phi$=0.944 in Fig.~\ref{fig_lowres_fragments}). As mentioned previously, the 
O III 5592 \AA\ line was shown to undergo RV variations consistent with a P$_C \sim$96.5d, 
$e\sim$0.8 orbit (Schweickhardt  2000).

\begin{figure} [!h, !t, !b]
\includegraphics[width=0.78\linewidth]{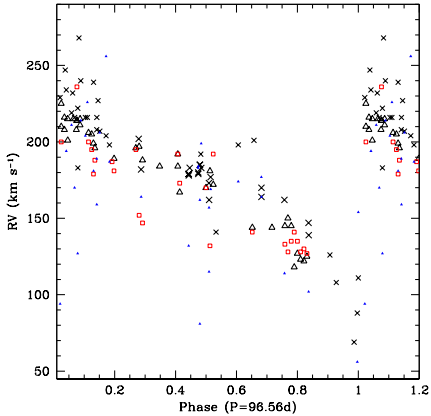}
\caption{RV measurements of the {\it Star C}  lines plotted against phase computed with P$_C$=96.56d.
Open triangles (LCO spectra) and crosses (FEROS spectra) correspond to the photospheric absorptions
of O III and He I; open squares (LCO) and filled-in triangles (FEROS) correspond to the 
weak C III 5696 emission line.
\label{fig_RVs_starC_all}}
\end{figure}

\begin{figure} [!h, !t, !b]
\includegraphics[width=0.98\linewidth]{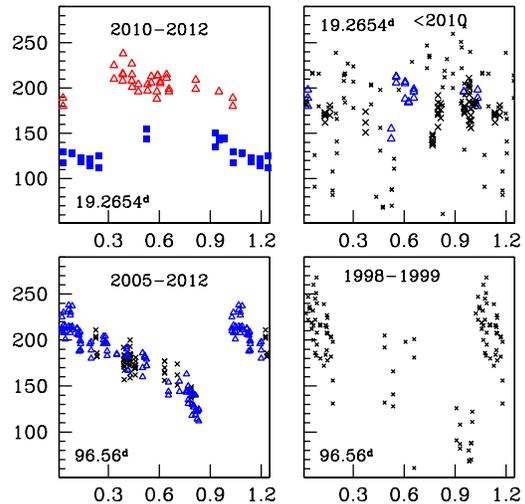}
\caption{RVs of the photospheric absorption lines O III 5592 and He I 4471, 5875 \AA\  plotted as
a function of phase computed with P=19.2654d (top) and 96.56d (bottom) for different
epochs of observation. The apparent modulation of the 2010--2012 data on the 19.2654d period
is due to a chance coincidence: open triangles and filled squares indicate, respectively,
 data with $\phi_C$=0.0-0.3 and 0.7--1.0, phase ranges at which maximum and minimum of the
96.56d modulation occur.
\label{fig_RVs_photospheric}}
\end{figure}

We selected  He I 4471 and 5875 \AA\ and O III 5592 \AA\ for measurement,  all of which 
are relatively uncontaminated by neighboring  lines and lie at reliable locations within the 
echelle orders. In addition, we measured the C III 5696 \AA\ emission line.
Tables ~\ref{table_LCO_starc} and ~\ref{table_FEROS_starC} list the RVs obtained by 
fitting Gaussians to the cores of the lines in the LCO and the FEROS spectra, respectively,
and Fig.~\ref{fig_RVs_photospheric} displays these values as a function of phase computed with the 
two periods: P$_{AB}$=19.2654 d (top panels) and P$_C$=96.56 d.
All RVs follow a clear periodic trend with P$_C$, as shown in the bottom panels, where
data from 2005--2012 (left) and from 1998-1999 (right) are plotted\footnote{The 1998-1999 data
are RVs measured on the same spectra that were used by Schweickhardt (2000).}.  Both of these RV sets
are combined in Fig.~\ref{fig_RVs_starC_all} illustrating a clear minimum at $\phi_C\sim$0
and maximum at $\phi_C\sim$0.05, consistent with Schweickhardt's conclusion that {\it Star C}
is a binary.  

It is a curious coincidence that P$_C$=5$\times$P$_{AB}$
which leaves the lingering suspicion that perhaps the data are really only modulated with
P$_{AB}$ but that the 1998-1999 activity levels of {\it Star A} prevented this modulation from being 
uncovered.\footnote{Schweickhardt searched for modulation on the P$_{AB}$ period in the 
OIII 5592 photospheric absorption RVs and convincingly showed that it was not present.}
Thus, we revisited this issue and plotted the 2010--2012 LCO RVs as a function of phase
using P$_{AB}$ (Fig.~\ref{fig_RVs_photospheric}, top left panel), finding a clear modulation
with this period.   However, upon closer inspection, we find that the 2010--2012 spectra
are clustered by chance in the two phase intervals $\phi_C$=0.0--0.3 and 0.7--1.0, which 
correspond, respectively, to times of maximum and minimum RVs.  This is shown by plotting with 
different symbols the data in each of these two phase intervales in the top
left panel of Fig.~\ref{fig_RVs_photospheric}.  This serves as a cautionary tale to be
heeded when finding modulations in data sets that incompletely sample the actual period. 
Furthermore, if we now plot the LCO data from observations prior to
2010 together with the 1998-1999 data,  the plot becomes a scatter diagram as found by Schweickhardt 
(2000) for the 1998-1999 observations.   Hence, we conclude that P$_C$ does indeed exist and can
only be associated with  {\it Star C} unless {\it Star A} or {\it Star B} undergo some kind of 
pulsation cycle on the P$_C$ timescale which also masks the RV variations on P$_{AB}$, something 
we do not consider very likely.

\begin{figure} [!h, !t, !b]
\plotone{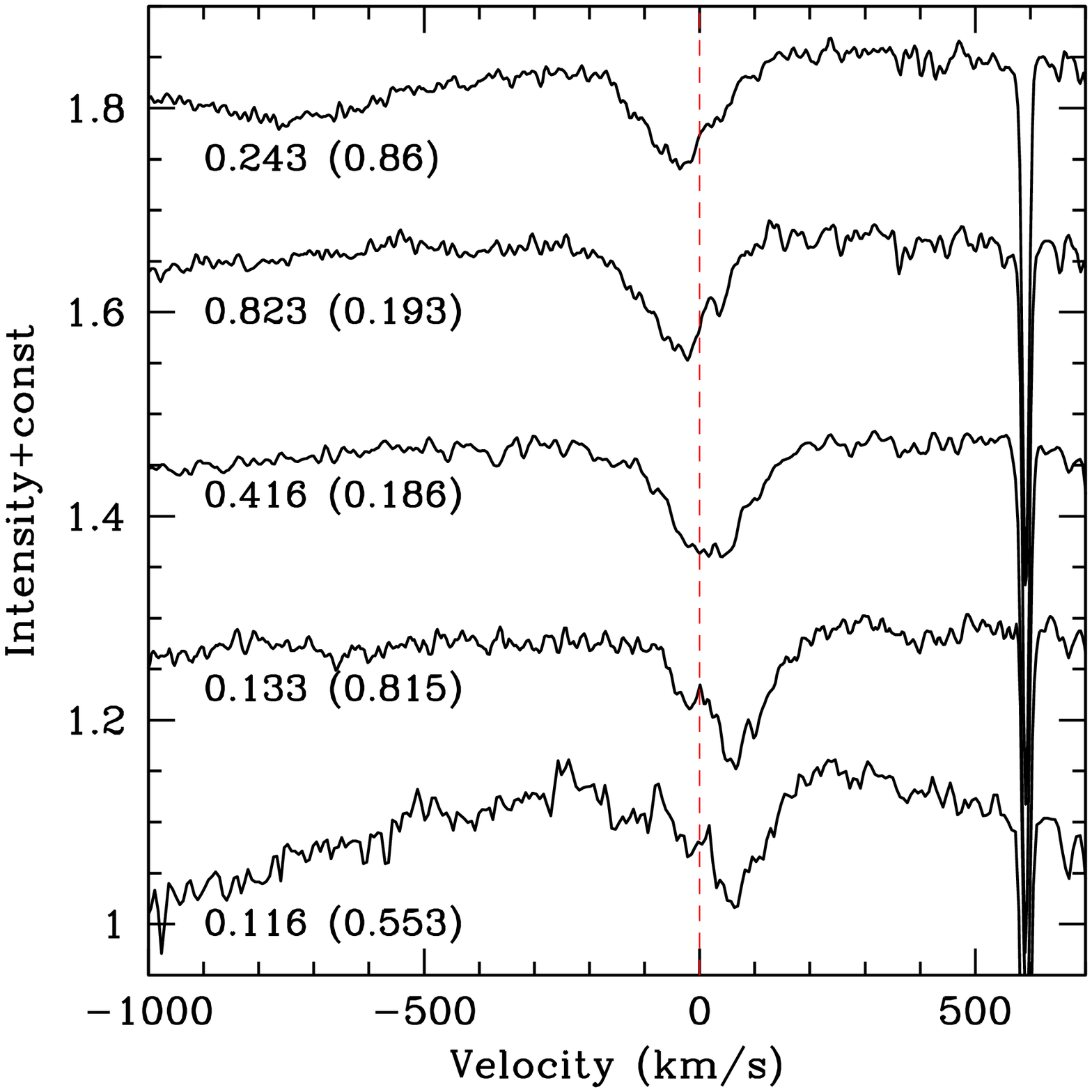}
\caption{The He I 5875 \AA\ photospheric absorption plotted on a velocity scale
corrected for +150 km/s SMC motion and displaced vertically in order of increasing
phase on the 96.56d period. The shift in the absorption  minimum is consistent with
orbital motion of the primary component of {\it Star C}, and there is marginal
evidence for the presence of a weaker absorption moving in anti-phase which, if
confirmed, would correspond to the secondary component. Listed are the phases for
the 96.56d period and, in parentheses, the phases corresponding to the 19.3d period.
\label{fig_montage_HeI5875}}
\end{figure}

Little can be said at this time regarding the nature of the secondary component of the
{\it Star C} system.  There is marginal evidence for the presence of weak absorptions that
may move in anti-phase with the primary lines as illustrated in Fig.~\ref{fig_montage_HeI5875}.
Significantly higher S/N ratio spectra are required for further analysis of this issue, however.

Finally, we note that the RVs of the weak C III~5696 \AA\ emission line follow the same trend
as the photospheric absorptions and thus it originates in {\it Star C}.  

\begin{deluxetable}{llrrrrr}
\tablecaption{LCO Radial velocity measurements Star C  \label{table_LCO_starc}}
\tablecolumns{7}
\tablewidth{0pt}
\tablehead{
\colhead{HJD} &\colhead{Phase}&\colhead{Phase}& \colhead{$\lambda$5592a} & \colhead{$\lambda$4471a} &\colhead{$\lambda$5875a}
& \colhead{$\lambda$5696e} \\
\colhead{$-$2450000} &\colhead{19.2654d}& \colhead{96.56d}&\colhead{km/s}& \colhead{km/s} & \colhead{km/s} &\colhead{km/s}
}
\startdata
    3920.861&     0.626&     0.350& \nodata&    191 &    192 & \nodata\\
    4342.731&     0.524&     0.719& \nodata&    139 &    150 & \nodata\\
    4670.799&     0.553&     0.116&    215 &    209 &    216 &    203 \\
    4671.823&     0.606&     0.127&    191 &    208 &    210 &    198 \\
    4672.799&     0.657&     0.137&    192 &    198 &    202 &    190 \\
    4774.800&     0.951&     0.193&    183 & \nodata&    183 &    174 \\
    5161.687&     0.033&     0.200&    186 &    177 &    168 &    169 \\
    5205.520&     0.308&     0.654&    148 &    138 &    134 &    135 \\
    5341.902&     0.388&     0.067&    245 &    227 &    250 & \nodata\\
    5342.889&     0.439&     0.077&    225 &    220 &    239 & \nodata\\
    5480.501&     0.582&     0.502&    175 &    159 &    149 &    159 \\
    5481.756&     0.647&     0.515&    170 &    170 &    172 &    121 \\
    5482.811&     0.702&     0.526&    167 &    160 &    160 &    180 \\
    5506.500&     0.931&     0.771&    118 &    137 &    122 &    115 \\
    5507.542&     0.985&     0.782&    133 &    132 &    131 &    122 \\
    5508.508&     0.035&     0.792&    127 &    105 &    117 &    128 \\
    5509.500&     0.087&     0.802&    128 &    114 &    116 &    122 \\
    5510.521&     0.140&     0.813&    127 &    110 &    106 &    115 \\
    5511.546&     0.193&     0.823&    110 &    109 &    101 &    117 \\
    5512.502&     0.243&     0.833&    111 &    112 &     99 &    114 \\
    5761.866&     0.186&     0.416&    186 &    171 &    171 &    177 \\
    5844.530&     0.477&     0.272&    189 &    185 &    193 &    184 \\
    5845.564&     0.531&     0.283&    187 &    186 &    193 &    141 \\
    5846.620&     0.586&     0.293&    172 &    176 &    184 &    136 \\
    6084.887&     0.953&     0.761&    174 &    156 &    154 &    144 \\
    6120.766&     0.815&     0.133&    193 &    204 &    214 &    184 \\
    6496.804&     0.334&     0.027&    216 &    213 &    228 &    203 \\
    6497.787&     0.385&     0.037&    233 &    211 &    219 & \nodata\\
    6498.833&     0.440&     0.048&    212 &    204 &    218 & \nodata\\
    6501.750&     0.591&     0.078&    232 &    216 &    216 &    238 \\
    6502.754&     0.643&     0.089&    239 &    217 &    213 & \nodata\\
    6533.737&     0.252&     0.409&    195 &    189 &    197 &    197 \\
\enddata
\end{deluxetable}
\section{Orbital solutions}

\subsection{The 19.3d WR/LBV system}

We employed the FOTEL\footnote{http://www.asu.cas.cz/~had/paifotel.pdf} program 
(\citet{2004PAICz..92...15H})  to fit the RV variations of the two Doppler-shifted emission lines of \ion{N}{5} 4944. 
Assigning the same weight to all the data points and fixing $i$=86$^\circ$ from Perrier et al. (2009)
the best solution gives  M$_A$=61$\pm$10 M$_\odot$, M$_B$=66$\pm$10 M$_\odot$, where the uncertainties 
refer to the goodness of fit. 
Our value for the eccentricity e=0.27$\pm$0.02 is slightly smaller that that derived by Perrier
et al. (2009) from the solution of the light curve (e$_{Perrier}$=0.314$\pm$0.007) but coincides with that
obtained by Schweickhardt (2000) from the RV curve solution of the 1998-1999 FEROS data, $e$=0.284.  Our 
argument of periastron $\omega$=134$^\circ\pm$4 coincides within the uncertainties with those of Perrier 
et al. (2009) and Schweickhardt (2000).

The FOTEL fit yields a period P=19.2656$\pm$0.0008d which coincides within the uncertainties with that
derived by Sterken \& Breysacher (1997), P$_{AB}$=19.2654 d.  Hence, we continue to adopt the 
Sterken \& Breysacher (1997) ephemeris in this paper. Our results do not change if the period is
fixed to this value. The time of periastron that is found by our fit, HJD2451424.97$\pm$0.25, 
corresponds to orbital phase $\phi_{peri}$=0.075.  

The parameters for the {\it Star A} and {\it Star B} system are listed in Table~\ref{table_RV_LC}.
The predicted RV curves are plotted in Fig.~\ref{fig_RVs_AandB},
displaced by -20 km s$^{-1}$, consistent with an expected systematic blue-shift
in \ion{N}{5} 4944 \AA\ due to occultation by the stellar core of the wind that is receding from the 
observer.
Our mass determination is consistent with the estimate derived in Foellmi et al. (2008), but it
is larger than that of Niemela (1988) and our M$_A$ is significantly smaller than the 90 M$_\odot$
estimated by \citet{2011AJ....142..191G}.

Niemela (1988) used the RV curves of the emission lines \ion{N}{5} 4603 \AA\ and \ion{N}{4} 4058 \AA\,
assumed to arise, respectively, from {\it Star A} and {\it Star B}. This explains the somewhat 
smaller masses because the \ion{N}{5} line is affected by the P Cygni absorption component of the 
neighboring \ion{N}{5} 4620 \AA\ line while the \ion{N}{4} emission contains a contribution also from
{\it Star A}.  

An answer to the question of whether the 90 M$_\odot$ deduced by \citet{2011AJ....142..191G} can be 
reconciled with our current results is less straightforward.
The spectroscopic mass in the CMFGEN model is based on the estimated luminosity and the proximity of the
star to the Eddington limit. For the 2009 model, {\it Star A} has  $L_{\hbox{\rm \small Edd}}/L=0.50$.
If we were to use 0.6 as an upper limit for the Eddington ratio, we would obtain M$_A$=75\,M$_\odot$,
in better agreement with that deduced from the radial velocity measurements. With
$L_{\hbox{\rm \small Edd}}/L\ge0.60$ it becomes difficult to construct a photospheric model since the
radiation force will potentially exceed gravity below the photosphere because of the additional opacity due 
to iron (as we will discuss below; see Figure 14). However, in this regime there are radiation instabilities that
need to be addressed, and clumping can potentially reduce the effective opacity (e.g., \citet{2001ApJ...549.1093S,
2001MNRAS.326..126S}).

Part of the discrepancy could also arise from the deduced luminosity since the deduced spectroscopic 
mass scales with $L$. If the error in $T_*$ is $\sim$3\% (2000\,K for $T_*=60,000$\,K)  the error in $L$ is
approximately 10\%. Given the additional uncertainties arising from the spectroscopic decomposition an
error in $L$ of 20\% is not unreasonable, and this translates to an error of almost 20\,M$_\odot$. 

Finally, we note that the uncertainty in the mass determinations quoted in Table 6 does not
include the uncertainties derived from possible contributions to the \ion{N}{5} 4944 \AA\ emission lines 
arising in, for example, the wind-wind collision region, a point that will be discussed in 
Section 5.3.

\subsection{The 96.56 d binary system}

The RVs of the O III 5592 and He I 4471, 5875 \AA\ photospheric absorptions 
were also analyzed with the FOTEL program (Hadrava 2004), from which  
the following orbital elements were derived: P$_C$=96.56d, $e$=0.82, $\omega$=252$^\circ$
and a time of periastron HJD2451183.40.  As shown in Table ~\ref{table_RV_LC_starC}, these values are 
consistent with those obtained by Schweickhardt (2000).  

The systemic  velocity that is derived from the average of the different
data sets (LCO, FEROS, 3 different lines) is V$_0^C$=164 $\pm$5 km s$^{-1}$.  This is
in contrast with the value we find for Stars {\it A} and {\it B}, V$_0^{AB}$=131 $\pm$3 
km s$^{-1}$.  However, as pointed out above, a systematic blue-shift in the NV 4944 \AA\
emission line is expected, so once the  -20 km s$^{-1}$ estimated shift is 
contemplated, the value of V$_0^{AB}$ is very similar to that of V$_0^C$, consistent with
both binary systems lying at the same location within the SMC.


\begin{deluxetable}{lllrrrr}
\tablecaption{FEROS Radial velocity measurements Star C  \label{table_FEROS_starC}}
\tablecolumns{7}
\tablewidth{0pt}
\tablehead{
\colhead{HJD} &\colhead{Phase}&\colhead{Phase}& \colhead{$\lambda$5592a} & \colhead{$\lambda$4471a} &\colhead{$\lambda$5875a} & \colhead{$\lambda$5696e} \\
\colhead{$-$2450000} &\colhead{19.2654d}& \colhead{96.5d}&\colhead{km/s}& \colhead{km/s} & \colhead{km/s} &\colhead{km/s}
}
\startdata
    1094.797&     0.935&     0.082&    251 &    183 &   \nodata &   \nodata \\
    1100.773&     0.245&     0.144&    214 &    208 &    189 &    189 \\
    1133.691&     0.954&     0.485&    205 &    192 &    132 &   \nodata \\
    1138.512&     0.204&     0.535&    166 &    141 &    128 &   \nodata \\
    1145.547&     0.569&     0.608&   \nodata & 198 &   \nodata &    174 \\
    1150.508&     0.827&     0.659&    181 &    201 &     61 &   \nodata \\
    1174.523&     0.073&     0.908&   \nodata & 126 &     91 &   \nodata \\
    1176.555&     0.179&     0.929&     82 &    108 &     87 &  \nodata\\
    1181.516&     0.436&     0.980&     80 &   \nodata&\nodata &     11 \\
    1183.523&     0.541&     0.001&    122:&    111:&   \nodata &    154 \\
    1185.562&     0.646&     0.022&   \nodata & 229 &   \nodata &    94:    \\
    1187.523&     0.748&     0.043&   \nodata & 234 &    236 &    194 \\
    1189.516&     0.852&     0.063&   \nodata & 232 &   \nodata &   \nodata \\
    1191.516&     0.955&     0.084&   \nodata & 268 &   \nodata &   \nodata \\
    1193.531&     0.060&     0.105&    217    & 216 &   \nodata &    204 \\
    1197.535&     0.268&     0.146&   \nodata & 227 &   \nodata &   \nodata \\
    1374.918&     0.475&     0.983&     68 &     69 &     44 &    -14 \\
    1375.922&     0.527&     0.994&     88 &     88 &     70 &     56 \\
    1379.902&     0.734&     0.035&    222 &    247 &   \nodata &   \nodata \\
    1381.902&     0.838&     0.056&    212 &    219 &   \nodata &    211 \\
    1382.852&     0.887&     0.066&    200 &    216 &    260 &    170 \\
    1383.855&     0.939&     0.076&    186 &    222 &    245 &    127 \\
    1384.809&     0.989&     0.086&    195 &    240 &    208 &    214 \\
    1386.891&     0.097&     0.107&    180 &    216 &   \nodata &    226 \\
    1388.840&     0.198&     0.128&    172 &    239 &   \nodata &    181 \\
    1389.883&     0.252&     0.138&    184 &    216 &    216 &    159 \\
    1390.828&     0.301&     0.148&    178 &    207 &    224 &    206 \\
    1392.828&     0.405&     0.169&    182 &    204 &    203 &    256 \\
    1393.871&     0.459&     0.180&    156 &    198 &    131 &    187 \\
    3538.894&     0.799&     0.394&    164 &    179 &    177 &    132    \\
    3541.890&     0.955&     0.425&    174 &    179 &    174 &    183 \\
    3561.877&     0.993&     0.632&    161 &    167 &    177 &    170 \\
    3576.795&     0.768&     0.787&    146 &    143 &    139 &    102 \\
    3635.520&     0.815&     0.395&    176 &    183 &    167 &   \nodata \\
    3638.763&     0.984&     0.428&    162 &    185 &    296 &    162 \\
    3641.662&     0.134&     0.458&    169 &    167 &    171 &    137 \\
    3665.527&     0.373&     0.706&    151 &    162 &    174 &    114 \\
    3715.789&     0.982&     0.226&    211 &    202 &    204 &   \nodata \\
    3716.539&     0.021&     0.234&    186 &    182 &    183 &    164 \\
    3731.539&     0.799&     0.389&    192 &    179 &    173 &   \nodata \\
    3735.551&     0.007&     0.431&    184 &    183 &    180 &    199 \\
    3738.539&     0.163&     0.462&    170 &    177 &    181 &    169 \\
\enddata
\end{deluxetable}

\subsection{Do the emission lines  reflect orbital motion ?}

The basic assumption that is made when using spectral lines to determine the orbital
parameters of a binary star is that these lines have a constant shape over the orbital
cycle.  In the case of an emission-line star, this translates into the assumption that
the distribution of emitting atoms in the material surrounding the star is symmetric and 
remains constant over the orbital cycle as viewed in the frame of the observer. If these 
conditions are not met, then the shape of the line profiles will change over orbital
phase, thus introducing a systematic deviation in the RV measurements with respect to 
those that yield orbital motion.

The issue of where in the HD\,5980 system the emission lines are formed has been a long-standing 
topic of debate
and has limited the efforts to determine the masses of {\it Star A} and {\it Star B}.   Moffat et al. (1998)
and Breysacher \& Fran\c{c}ois (2000) speculated that the bulk of the emission lines arises not in the stellar
winds, but in the shock cone region where the two winds collide (henceforth referred to as the wind-wind
collision, WWC).  If this were the case,
then the RV variations would describe the velocity structure along the shock cone walls, projected
along the line of sight to the observer, and their use for deducing the stellar masses would be very limited.  

Although the WWC shock walls may indeed give rise to emission lines, there are additional effects that may 
also distort the emission-line profiles.  For example, the  wind region belonging to the star with the
weaker wind that is enclosed by the WWC constitutes a ``hole'' in the spherical distribution of the stronger 
wind star. It has been shown that this effect results in weaker emission at particular wavelength
intervals which change with orbital phase (Georgiev and Koenigsberger 2004).
Another interaction effect is the irradiation of {\it Star A}'s wind by {\it Star B}, which may alter the 
ionization and velocity structure (\citet{1997ApJ...475..786G}) on the affected hemisphere.  One-dimensional model 
calculations of the P V line profile variations in {\it FUSE} spectra led to the conclusion that the wind 
structure of {\it Star A} is different on the hemisphere facing the companion from that of the opposite 
hemisphere (\citet{2006AJ....132.1527K}), supporting this scenario. 

Although, a proper analysis of irradiation, WWC and asymmetric wind structures requires the use of  
2 or 3-dimensional non-LTE radiative tranfer computations, the dominant effect of these interaction effects 
is expected to be present in the outer wind regions.  Hence, as we shall argue below, emission lines
which are formed in the innermost wind regions, such as NV 4944 \AA, are likely to be much less affected.

\begin{figure} [!h, !t, !b]
 \plotone{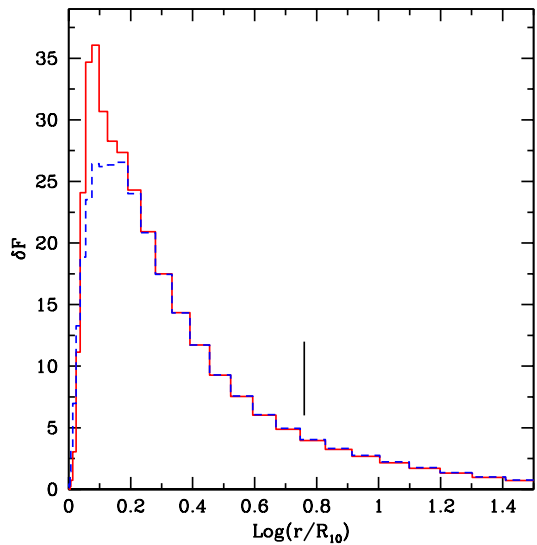}   
\caption{Shown (red, solid) is the radial contribution to the flux (N V 4994 \AA\ line and underlying continuum)  
measured over a band extending from 4942.3 to 4945.3 \AA. Also shown (blue, broken) is the radial contribution 
to the continuum flux near the N V 4994 line.  These data were obtained from the CMFGEN model constructed for 
{\it Star A} for the 2009 epoch (\citet{2011AJ....142..191G}). The radius variable is given in units of 
R$_{10}$=19.3 R$_\odot$, 
which corresponds to the location where the continuum optical depth $\tau_{Ross}$=10. Adopting the orbital 
elements listed in Table 6, the separation between Stars A and B at periastron is log(R$_{per}$/R$_{10}$)=0.76 
(indicated by vertical line), which is outside of the wind region where this \ion{N}{5} line primarily forms.
\label{fig_hillier_nv} }
\end{figure}

Weak emission lines such as \ion{N}{5} 4944 \AA\ are formed in a small volume, one that is comparable to 
that of the continuum forming region, as we illustrate in Fig.~\ref{fig_hillier_nv} which shows 
the extent of the \ion{N}{5} 4944 \AA\ forming region computed by the CMFGEN model for {\it Star A} 
(\citet{2011AJ....142..191G}).  The bulk of this emission arises at log(r/R$_{10}) \leq$0.3,
where R$_{10}$ is the radius at which the continuum optical depth $\tau$ is 10.  Adopting 
R$_{10}$=19.3 R$_\odot$ from \citet{2011AJ....142..191G}, we find that the bulk of the \ion{N}{5} 4944 \AA\
is formed at $<$40 R$_\odot$, which is significantly smaller than the orbital separation, even 
at periastron passage (r$^{peri}\sim$110 R$_\odot$).  This supports the assumption that
the RV curve of this \ion{N}{5} line is likely to be unperturbed by interaction effects.  

A further test is to compare the CMFGEN model line profiles with the observations, for which 
we synthesized spectra of the HD\,5980 system under the following conditions:  a) the relative continuum
brightnesses of {\it Star A}, {\it B} and {\it C} in the visual spectral region are, respectively,
40\%, 30\% and 30\%;  b) the emission-line spectrum of {\it Star B} is approximated with the CMFGEN 
emission-line spectrum of {\it Star A} computed by \citet{2011AJ....142..191G} for the 2009 epoch, scaled 
to the relative luminosity of {\it Star B}; c) the spectrum of {\it Star C} is represented by the 
appropriate CMFGEN model (\citet{2011AJ....142..191G}). 

The CMFGEN model spectra for {\it Star A} and {\it Star B} were Doppler-shifted to the
radial velocities measured on the \ion{N}{5} 4944 \AA\ line in the FEROS spectra
at orbital phases $\phi$=0.162 and 0.799.  These were then combined with each other and
with the spectrum of {\it Star C} to create the synthetic spectrum of the triple system
at these two orbital phases.

\begin{figure} [!h, !t, !b]
\plotone{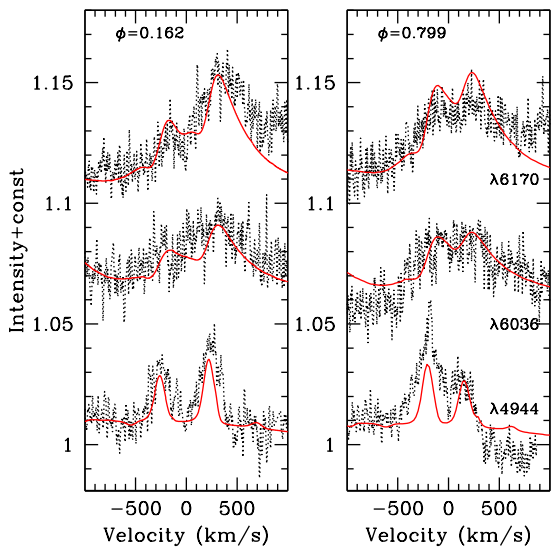}
\caption{Comparison of  FEROS spectra (dots) obtained at two orbital phases
with the synthetic spectrum constructed for the HD\,5980 system using the CMFGEN model computed
for {\it Star A} and {\it Star C} by \citet{2011AJ....142..191G} and assuming that {\it Star B}'s spectrum is
similar to that of {\it Star A}.  The velocity scale is centered on the laboratory wavelength of
\ion{N}{5} 4944 \AA\ and \ion{He}{2} 6036 and 6170 \AA.
\label{compare_two_panels_CMFGEN1} }
\end{figure}

\begin{figure} [!h, !t, !b]
\plotone{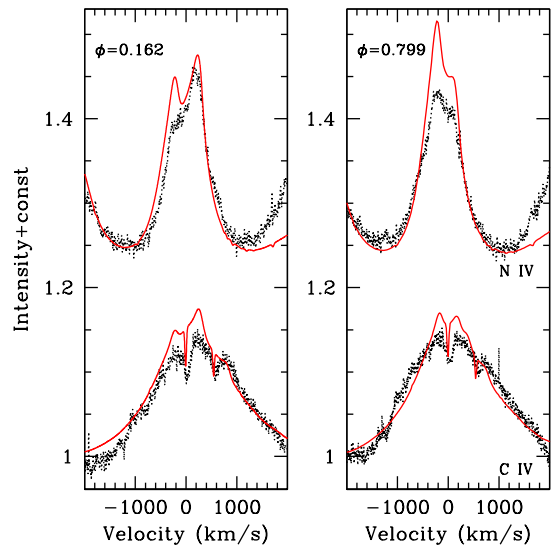}
\caption{Same as Fig. ~\ref{compare_two_panels_CMFGEN1} but here the velocity
scale is centered on the laboratory wavelength of \ion{C}{4} 5801 \AA\ (bottom)  and \ion{N}{4} 4058 \AA\ (top).
\label{compare_two_panels_CMFGEN2} }
\end{figure}

Figs.~\ref{compare_two_panels_CMFGEN1} and \ref{compare_two_panels_CMFGEN2} 
present a selection of emission-line profiles in the spectra obtained
in the year 2006 at orbital phases $\phi$=0.162 (left panel) and $\phi$=0.799
(right panel) compared with the synthetic spectrum computed for the corresponding phases.
Other than the discrepancies in line strengths which are to be expected given the
uncertainties in the input parameters for the CMFGEN model,  the synthetic triple-star 
spectrum reproduces quite well the observations.  Hence, there is no indication that the 
lines shown in these figures are affected in any significant way by the interaction effects 
in the binary system.  

It is also important to note that the \ion{N}{5} 4944 \AA\ emission at orbital phase $\phi$=0.162
consist of two {\em separate} lines, both of which reach the local continuum level; i.e.,
they are not really  ``double peaked'' emission lines, a shape predicted from  models
of emission arising in the WWC shock cone walls (see, for example, \citet{2003MNRAS.346..773H}).

Finally, we note that although winds with similar momentum ratios lead to a WWC region in which
the phase-dependence of emission line profiles is similar to that of two distinct wind sources,
to the extent of our knowledge, no calculation of the intensity of such emission lines has been
performed.  Thus, it is not possible to assess the relative importance of this emission compared
to that which arises from the inner wind regions of the two stars.



Hence, we conclude that, given the information for HD\,5980 that is available at this time and 
given the lack of  models that predict the strength of the emission lines expected to arise in a
wind-wind collision region, the assumption that the \ion{N}{5} 4944 \AA\ emission line arises primarily in
the inner stellar winds of the stars is a sound one.  

Clearly, however, this conclusion is based on the premise that {\it Star B} possesses an 
emission-line spectrum that arises in a high-ionization stellar wind typical of WNE-type  stars. 
This premise is based on observations obtained prior to and during the eruptive state 
(Breysacher et al.1982; Niemela 1988; Koenigsberger 2004). If {\it Star B} were to be a non-WR star, 
then the WNE-type spectrum that has been observed for $>$60 years would have to arise entirely in 
{\it Star A} and, potentially, in the WWC region, and the conclusions of this paper would have to be
re-assessed.  It is important to note, however, that the wind momentum ratio under these circumstances
would most likely be such that the WWC shock cone would be folded around {\it Star B} and hence, a
highly blue-shifted emission component would be observed in lines of He I and \ion{He}{2} near the conjunction 
when this star is between us and its companion\footnote{See illustrations of a similar geometries in Figs. 
20-22 of Koenigsberger 2004}, something that is not observed.

\begin{deluxetable}{lrrr}
\tablecaption{Orbital solutions for Stars A and B \label{table_RV_LC}}
\tablecolumns{4}
\tablewidth{0pt}
\tablehead{
\colhead{} & \multicolumn{2}{c}{NV 4944 RVs}   &  \colhead{}      \\
\colhead{Element} &\colhead{Star A} &\colhead{Star B}  &\colhead{System A+B}
}
\startdata
M sin$^3$ i (M$_\odot$)& 61 (10)              &  66 (10)     & 127 (14)    \\
a~sin~i (R$_\odot$)    &   78 (3)             &   73 (3)     &  151 (4)\\
K (km s$^{-1}$)        & 214 (6)              &  200 (6)     & \nodata  \\
e                      & \nodata              & \nodata      & 0.27 (0.02)  \\
$\omega_{per}$ (deg)   &  \nodata             & \nodata      & 134 (4)  \\
V$_0$ (km/s)           &  \nodata             & \nodata      & 131 (3) \\
P$_{calc}$ (days)      & \nodata              & \nodata      & 19.2656 (0.0009) \\
T$_{peri}$ (HJD)       & \nodata              & \nodata      & 2451424.97 (0.25) \\
$i^\circ$              & \nodata              & \nodata      & 86 (fixed) \\
T$_0$ (HJD)            & \nodata              & \nodata      & 2443158.71 (fixed) \\
P$_{AB}$ (days)        & \nodata              & \nodata      & 19.2654 (fixed)    \\
\enddata
\end{deluxetable}

\begin{deluxetable}{lrrr}
\tablecaption{Orbital solution Star C \label{table_RV_LC_starC}}
\tablecolumns{3}
\tablewidth{0pt}
\tablehead{
\colhead{Element} &\colhead{Current analysis}& \colhead{Schweickhardt (2000)} }
\startdata
P$_C$ (days)           &  96.56 (0.01)     &    96.5         \\
T$_{peri}$ (HJD)       &2451183.40 (0.22)  & 2451183.3         \\
$e$                    &  0.815 (0.020)    &    0.82         \\
$\omega$ (deg)         &  252   (3.3)      &    248          \\
K (km/s)               &  81 (4)           &     76     \\
\enddata
\end{deluxetable}

\section{Evolutionary status}

In this section we will examine the evolutionary scenarios that lead to a system such as 
the 19.3d binary in HD\,5980.  

A preliminary examination (Koenigsberger 2004) indicated that
the luminosity and effective temperature of {\it Star A} place it on the H-R 
Diagram along the evolutionary track of a 120 M$_\odot$  zero-age main sequence (ZAMS) star 
at an age of $\sim$3 Myr.  This is  very close to the age of the NGC 346 cluster (\citet{2006A&A...456.1131M}),
which would indicate that such a massive star is near the final evolutionary phases
leading up to a supernova event. A similar result is reached if we use the most recently available 
grid of  models for the SMC metallicity (\citet{2013A&A...558A.103G}). Specifically, {\it Star A}'s luminosity
corresponds to that of a non-rotating 120 M$_\odot$ ZAMS model near the end of core helium
burning at an age of 2.952 Myr or to that of a 85 M$_\odot$ model with initial rotation velocity
of 400 km s$^{-1}$ at an age of 3.63 Myr.  However, both of these models predict a large increase
in radius prior to these ages.  Since the current Roche-lobe radius is only $\sim$ 57 R$_\odot$,
the interaction with the companion would have modified the subsequent evolution from that predicted
by the single-star models.  Thus, HD\,5980 needs to be analyzed within the context of  binary
evolution or within scenarios in which the large radius increase in post-main sequence
stages can be avoided.

The classical binary scenarios involve the assumption that a star evolves
as a single star until its radius reaches some critical value, assumed to be the
Roche radius. At this point in the evolution, the nature of the close companion
becomes a crucial element of the analysis because it determines whether conservative or  
non-conservative mass transfer takes place.   Our results indicate
that the mass of {\it Star B} is close to that of {\it Star A} and that it, too, has a very
large surface He-abundance enrichment. This excludes the conservative mass transfer scenario
which requires rather similar masses initially, and thus leads to large mass ratios after
the RLO  phase (\citet{1999A&A...350..148W}).   Non-conservative interaction, on the
other hand, as expected for relatively large initial mass ratios (\citet{2005A&A...435..247P}),
could leave two similar mass objects after the envelope of the initially more massive
star is stripped. However, the initially less massive star would still be relatively
unevolved at this time, in contrast with the rather evolved state of both stars in HD\,5980.
Hence, the classical binary evolution scenarios cannot readily explain the parameters of
HD\,5980.

The question then remains whether it is possible to avoid the large radius increase in
the post-MS evolutionary stages, thus avoiding the problem of mass transfer in the system,
but at the same time enhancing the surface chemical composition.

Rotational mixing
brings the helium  and nitrogen produced in the core efficiently to the stellar surface,
thereby enriching the visible surface chemical abundances without the need of stripping
large amounts of mass from these outer layers. Since this rapid rotation greatly reduces
the chemical gradients present in slower rotating stars, these stars undergo what is
referred to as quasi-chemically homogeneous evolution (QCHE). An important characteristic of
these models is that {\em both} L and T increase with time, and the star remains
relatively compact thus avoiding a RLO phase.  According to \citet{2011A&A...530A.116B} , stars with 
rotational velocities above $\sim$300 km s$^{-1}$ are at the threshold for QCHE 
at SMC metallicity.

\begin{figure} [!h, !t, !b]
\includegraphics[width=0.70\columnwidth,angle=-90]{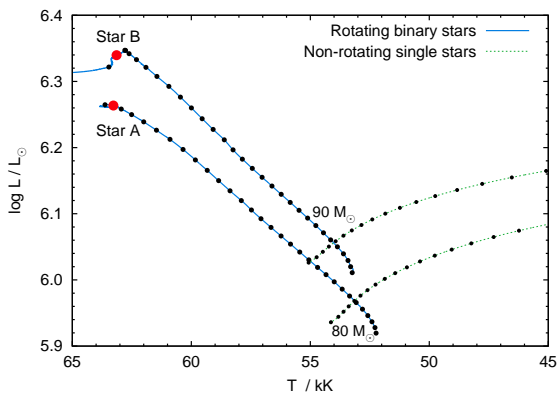}
\caption{Hertzprung-Russell diagram (HRD) of our binary model with an initial orbital period
of 12d, initial surface otational velocity v$_{ini}$=500 km s$^{-1}$, and inital masses of
90 M$_\odot$ and 80 M$_\odot$, compared for reference to the evolution of same mass non-rotating
stars.  Black dots are shown every 0.1 Myr.  The red  circles indicate the position of the two
HD\,5980 binary components after $\sim$3.1 Myrs, which is when their surface properties agree
best with those of {\it Star A} and {\it Star B}.
\label{HRD_bonn}}
\end{figure}

We computed a model binary system with initial masses 90 M$_\odot$ (Primary) and 
80 M$_\odot$ (Secondary), rotational velocity v$_{ini}$= 500 km s$^{-1}$  and
orbital period of 12d using the Binary Evolutionary Code (\citet{2000ApJ...528..368H, 2005A&A...435..247P,
2006A&A...460..199Y}). For the early evolutionary stage, the prescription of \citet{2001A&A...369..574V} for the
mass-loss rate was adopted and for the WR stage we used the prescription of   \citet{1995A&A...299..151H}, 
with the correction  of \citet{2005A&A...442..587V} to account for the SMC metallicity.  Because
of the angular momentum loss due to the stellar winds, the orbital separation grows with time.
After $\sim$3.1 Myr, the binary has an orbital period of 19.2d and the stars have each $M$=67 M$_\odot$,
log(L/L$_\odot$)$\sim$6.3, highly enhanced surface He-abundances, and current surface temperatures
of 63 kK,  similar to the properties of {\it Star A} and {\it Star B}.  More importantly, the 
evolution of these QCHE model stars places them on the HRD at systematically higher temperatures and 
luminosities, as shown in Fig.~\ref{HRD_bonn}, without any large radius increases that would lead to 
mass transfer.

\begin{deluxetable}{lllcl}
\tablewidth{0pt}
\tablecaption{Binary evolution model\tablenotemark{a} }
\tablecolumns{5}
\tablehead{\colhead{Parameter} &\colhead{Primary} & \colhead{Secondary}&\colhead{{\it Star A}}  &\colhead{Notes\tablenotemark{b}}}
\startdata
 Age (Myr)                                 & 3.1&3.1        &$\sim$3  &  1        \\
 M$_{ZAMS}$ (M$_\odot$)                    & 90 & 80        &\nodata  & \nodata   \\
 M (M$_\odot$)                             & 67 & 67        &51--71   &this paper \\
 P$_{orb}$ d                               &19.2&19.2       &19.26    &           \\
 log(L/L$_\odot$)                          & 6.34 &  6.26   & 6.39    & 3         \\
 $Y$                                       & 0.92 & 0.84    & 0.80    & 3,4       \\
 $T_{phot}$ (kK)                           & 63.0 & 63.0    & 60      & 3         \\
 $v_\mathrm{surf}$                         & 195 & 257      & 250:    & 2         \\
 R$_{phot}$ (R$_\odot$)                    & 12.5 & 11.3    &19--24   & 5         \\
 $\dot{M}$(10$^{-5}$M$_{\odot}$ yr$^{-1}$) & 5.5 & 3.1      & 3.6     & 3         \\
\enddata
\tablenotetext{a)} {The computed model is with initial orbital period of 12 d, surface rotational
velocity v$_{ini}$= 500 km s$^{-1}$, and metallicity Z=0.0021. v$_{surf}$ is the surface rotation velocity.}
\tablenotetext{b)}{References for Columns 2 and 3: $^1$Mokiem et al. 2006; $^2$Georgiev \& Koenigsberger 2004;
$^3$Georgiev et al. 2011; $^4$Koenigsberger et al. 1998; $^5$Perrier et al. 2009}
\end{deluxetable}

The other parameters of the QCHE model stars are listed in Table 8, along with the corresponding
values for {\it Star A}.  Since the Secondary is the originally less massive member of the binary 
system, it evolves slower than the companion and thus corresponds to {\it Star A}, so the comparison 
to be made is between {\it Star A}'s observationally-derived parameters and those of the secondary 
in the QCHE model.  The mass-loss rate of {\it Star A} in 2009 is listed by \citet{2011AJ....142..191G} as
$\dot{M}/\sqrt(f)$=23$\times$10$^{-5}$ M$_\odot$ yr$^{-1}$, and they adopt a wind ``clumping'' factor 
$f$=0.025.  Hence, $\dot{M}_A$=3.6$\times$10$^{-5}$ M$_\odot$ yr$^{-1}$, which is very similar to
that of the Secondary in the QCHE model.   

Although there is no direct means of determining the surface rotation
velocity of the WR-type stars in HD\,5980, Georgiev \& Koenigsberger (2004) speculated that the
low-amplitude modulation of {\it Star C}'s photospheric absorptions that was reported in early
observational data by Breysacher et al. (1982)  could be caused by a much larger amplitude Doppler 
shift of underlying broad lines which, due to their width, would not be detectable in the HD\,5980 spectrum.  
Under this assumption, they found that the possible absorption lines would correspond to {\it Star A} 
rotation speeds of  $v~sini \sim$200--250 km s$^{-1}$, which is of the order of the rotation predicted for 
the Secondary in the QCHE model.

The final point to note concerns the predicted radii of the QCHE model stars,
R$_{Prim}$=12.5 R$_\odot$ and R$_{Sec}$=11.3 R$_\odot$.  The analysis of the optical eclipse light
curve of HD\,5980 (Perrier et al. 2009) yielded relative radii of $\rho_A/a$=0.158 and $\rho_B/a$=0.108,
where $a$ is the semi-major axis of the orbit. From Table 6, $a$=151 R$_\odot$ which implies that the occulting 
disks  are R$_B \sim$16 R$_\odot$ and R$_A \sim$24 R$_\odot$.  The latter is $\sim$2 time larger than 
predicted by the QCHE model.   However, it must be considered that: 1) the derived radius for the occulting 
disk in WR stars depends on the wavelength of observations, with smaller wavelengths yielding smaller radii 
(\citet{1984ApJ...281..774C}),  so a small correction may be viable if UV light curves are analyzed; and 
2) the light curve solution of Perrier et al. (2009) is based on the assumption that only {\it Star B} 
possesses an extended optically thick wind, so it remains to be determined whether assuming that both 
stars have such a region would result in a smaller radius for {\it Star A}.

We thus conclude that a plausible scenario for the evolutionary state of the 19.3d binary in 
HD~5980 is one in which both stars have followed QCHE pathways.  

\begin{figure} [!h, !t, !b]
\includegraphics[width=0.70\columnwidth,angle=-90]{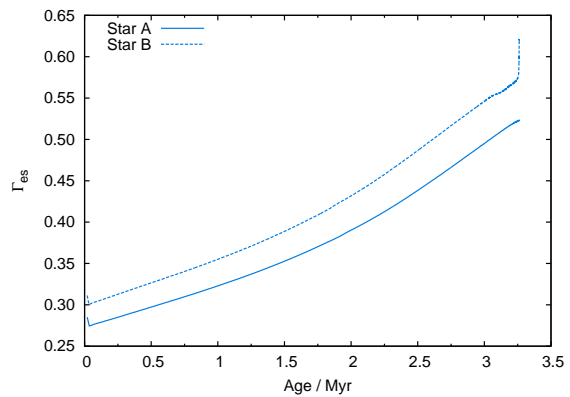}
\caption{Value of the electron scattering Eddington factor over time since arriving
at the Main Sequence showing its systematic increase.
\label{fig_gamma_es_bonn}}
\end{figure}

\begin{figure}
\includegraphics[width=0.70\columnwidth,angle=-90]{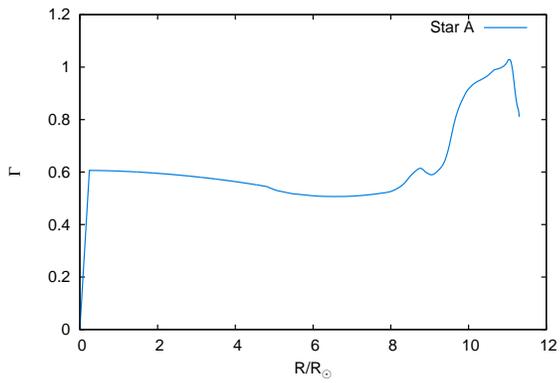}
\caption{Value of the total Eddington factor, calculated using the total opacity (not just
that due to electron scattering) and the radiative flux in {\it Star A} as a function of
distance from stellar core.  Note the super-Eddington values very close to the surface that
are caused by the Fe-opacity ``bump''.
\label{fig_gemma_tot_bonn}}
\end{figure}

Furthermore, the internal luminosity structure of the QCHE models is such that therein may 
reside the mechanism that triggered the sudden eruption observed in 1994.  The increasing 
luminosity over time causes an increase in the Eddington electron scattering $\Gamma$-factor 
($\Gamma_{es}$),  thus lowering the effective gravity at the surface.  Fig.~\ref{fig_gamma_es_bonn} 
shows how the $\Gamma_{es}$-factor grows beyond 0.5 as the age of the star approaches that  
of {\it Star A}.  This is consistent with the value of $\Gamma_{es}$ found by \citet{2011AJ....142..191G} 
from the spectrum of 2009.  More important, however, is the fact that the {\em total}  
$\Gamma$ (i.e., that computed with the total opacity, not only the electron scattering opacity) 
becomes larger than unity near the surface layers (Fig.~\ref{fig_gemma_tot_bonn}).  One possible 
mechanism driving the eruption observed in 1994 might be related to this super-Eddington layer which might
turn unstable (\citet{1994PASP..106.1025H,1996ApJ...468..842S}). We speculate that in the case of HD\,5980, the
instability could have been triggered by the pertubations that arise due to its excentric orbit.
A discussion of the potential role of the companion in the case of $\eta$ Car has also been discussed
by  \citet{2012ASSL..384...43D}.

The question that then arises is whether the same mechanism could be active in other LBVs.
An objection to the QCHE scenario is that in order
for stars to follow this path, the rotation rate needs to remain high throughout their
evolution, a condition that is difficult to satisfy due to the expected loss of angular
momentum through stellar winds and tidal interactions.  To avoid this problem, de \citet{2009A&A...497..243D} 
suggested that very short-period, tidally-locked binary systems could sustain the very
rapid rotation velocities that are required.  However, none of the known LBV binaries are in
orbits that are close enough to meet this criterion, even after considering orbital evolution.

However, the important point to note is that for QCH evolution to take place, the  star
must be able to efficiently mix the nuclear processed material up to the surface layers,
and rapid rotation may not be the only mechanism that can produce this effect. 
For example, though speculative at
this time, an alternative to rapid rotation for efficient mixing might be the time-dependent velocity 
field that is driven by the tidal interactions in an asynchronously rotating binary star (\citet{EAS:9185257}).
With this mechanism, one could start out with a relatively wider binary than is required 
by the fast-rotation models and, as long as the stellar rotation remains asynchronous, the induced 
velocity field could contribute toward the mixing processes.  Whether the efficiency with which the 
mixing occurs is sufficient for QCHE to be attained, however, remains to be proven.

Little can be said at this time about the evolutionary status of the {\it Star C} binary except
for noting that such highly eccentric binary systems are rarely found in orbits with periods 
shorter than 100 d, which suggests that it may be gravitationally bound to the 
{\it Star A} + {\it Star B} pair.  The 5:1 resonance in the orbital periods of the two systems
is further evidence in support of this hypothesis.  In triple hierarchical systems, the outer
period tends to be a multiple of the inner binary period, as in $\sigma$ Ori (Aa+Ab) + B 
(Sim\'on-D\'{\i}az et al. 2011).  If a similar condition were to hold in HD\,5980, then both
P$_{AB}$ and P$_C$ would be sub-multiples of the outer period and hence, would be related 
to each other, as observed. Thus, we speculate that the {\it Star C} components have a similar age 
as {\it Star A} and {\it Star B}.

\section{Conclusions}

Due to their short lifetime, LBVs are quite rare and most of them are apparently single objects,
impeding a direct determination of their current mass.  Determining basic parameters for the LBVs
undergoing major eruptions is even more challenging, since the majority are detected only 
in distant galaxies and little or no information exists on the properties of the star prior to the 
eruptive event. 

In this context, the importance of HD\,5980 for understanding massive star evolution cannot be overstated.
It provides the closest and clearest example of mass-loss  phenomena during the short-lived LBV
and WR phases in low-metallicity star forming regions such as those where the most distant LBV events are
being observed and which, in addition, host the brightest gamma-ray events.  Furthermore, 
massive objects such as those in HD\,5980 are believed to be the progenitors of some of these events.  

The following are the conclusions of this paper:

1.- The masses in the 19.3 d {\it Star A}+{\it Star B} system have been constrained under
the premise that both stars currently possess a WNE-type spectrum.  The largest uncertainties in this
determination derive from: a) the fact that the spectrum of {\it Star B} has not been unambiguously 
isolated from that of {\it Star A}; and b) the lack of detailed models predicting the strength of the
emission lines that can arise in the wind-wind collision region.  

2.-  A plausible scenario for the evolutionary state of the 19.3d binary in HD~5980 is one in which 
both stars have followed a quasi-chemically homogeneous evolutionary (QCHE) path.  We speculate that 
the high mixing 
efficiency that is required may be provided, at least in part, by the non-uniform differential rotation 
structure that is driven by the tidal forces in an asynchrounously rotating binary star.
HD\,5980 may thus be the first binary system which provides evidence that quasi-chemically 
homogeneous evolution can indeed occur, since it possesses the two extreme properties which are
required for this scenario, i.e. a very high mass of its components, and a very low metallicity. If
confirmed this would imply that both stars in HD\,5980 can be considered as candidates for gamma-ray
burst progenitors (Yoon et al. 2006) or even pair instability supernovae (\citet{2007A&A...475L..19L}).

3.- The instability that caused the 1994 eruption may be related to the super-Eddington layer which 
is present just below the photosphere in the  QCHE models.

4.-  A 96.5 d period is confirmed for the radial velocity variations of
  the photospheric absorptions arising in the third component, {\it Star C},
  indicating that this star is also a binary. We point out that this period
  is in a 5:1 ratio with the period of the {\it Star A} + {\it Star B} binary,
  suggesting that these four stars may constitute a hierarchical quadruple
  system.

\begin{acknowledgements}
We dedicate this paper to the memory of our friends and colleagues Virpi Niemela and 
Leonid Georgiev.

We thank Sylvia Ekstrom for providing the SMC evolutionary models in advance of publication and Ulises Amaya
and Francisco Ru\'{\i}z Salazar  for computer support.  GK acknowledges UNAM/PAPIIT grant
IN 105313 and CONACYT grant 129343. RHB acknowledges support from FONDECYT Regular Project 1140076.  
DJH acknowledges support from Grant HST-GO-11623.01-A.  The HST is operated by STScI under contract 
with AURA.  We thank Eric Hsiao (Carnegie Supernova Project) and Josh Simon for the
MIKE spectrum of June 6, 2012; and Ricardo Covarrubias for the August 30,
2007 MIKE spectrum taken during Clay engineering time. We are indebted to
the Carnegie TAC for generous observing time allocation at the du Pont
telescope, and to the authorities of Las Campanas for offering us the
opportunity to observe at du Pont during engineering time in July 2013.
\end{acknowledgements}

\clearpage

\bibliographystyle{apj}
\bibliography{antonio_rivera_HD5980.bib} 

\end{document}